\documentclass[%
 reprint,
superscriptaddress,
 amsmath,amssymb,
 aps,
floatfix, 
]{revtex4-2}

\usepackage{graphicx}
\usepackage{dcolumn}
\usepackage{bm}
\usepackage{xcolor}
\usepackage{comment}
\usepackage{physics}
\usepackage{braket}
\usepackage{mathtools}
\usepackage[hidelinks]{hyperref}

\newcommand{\cancel}[1]{}

\begin{document}

\preprint{APS/123-QED}

\title{A Pfaffian quantum Hall state of ultracold bosons}

\affiliation{Department of Physics, Harvard University, Cambridge, MA 02138, USA}
\affiliation{Department of Physics and Arnold Sommerfeld Center for Theoretical Physics (ASC),
Ludwig-Maximilians-Universit\"{a}t M\"{u}nchen, Theresienstr. 37, M\"{u}nchen D-80333, Germany}
\affiliation{Munich Center for Quantum Science and Technology (MCQST), Schellingstr. 4, D-80799 M\"{u}nchen, Germany}
\affiliation{Institut f\"{u}r Theoretische Physik und Astrophysik and W\"{u}rzburg-Dresden Cluster of Excellence ct.qmat,
Julius-Maximilians-Universit\"{a}t, 97074 W\"{u}rzburg, Germany}

\author{Joyce~Kwan$^{1,\dagger,*}$}
\author{Perrin~Segura$^{1,*}$}
\author{Yanfei~Li$^{1}$}
\author{Tizian~Blatz$^{2,3}$}
\author{Annie~Zhi$^{1}$}
\author{Brice~Bakkali-Hassani$^{1,\ddagger}$}
\author{Annabelle~Bohrdt$^{2,3}$}
\author{Martin~Greiter$^{4}$}
\author{Fabian~Grusdt$^{2,3}$}
\author{Markus~Greiner$^{1,\S}$}

\date{\today}

\begin{abstract}
Fractional quantum Hall states are a cornerstone of topological physics, hosting fractionally charged quasiparticles with exotic statistics that promise to enable topologically protected quantum information processing. Among these, the Pfaffian state introduced by Moore and Read implements a p-wave pairing structure that supports excitations with non-Abelian exchange statistics~\cite{moore_nonabelions_1991,greiter_paired_1991}. Despite extensive study in electronic systems, direct access to its pairing structure has remained limited. Here we realize a three-particle bosonic Pfaffian state of ultracold $^{87}\mathrm{Rb}$ atoms in an optical lattice subject to a Floquet-engineered synthetic magnetic field. Using a Bayesian-optimized adiabatic protocol~\cite{blatz_bayesian_2024}, we prepare a state exhibiting Pfaffian pairing correlations. Site-resolved measurements of multi-point density correlations reveal a pronounced suppression of short-range three-body coincidences, reflecting the underlying pairing structure. We further probe the state's transport response through Hall drift measurements~\cite{repellin_fractional_2020}. Our results establish a bottom-up approach to engineering non-Abelian topological order and lay the groundwork for future explorations of anyonic braiding in synthetic matter.
\end{abstract}

\maketitle

\section*{\label{sec:level1}Introduction}
The fractional quantum Hall (FQH) effect hosts a plethora of topologically ordered phases characterized by fractionally quantized Hall conductance and quasiparticles with fractional charge and statistics. They arise when strongly interacting particles confined to two dimensions in a magnetic field form incompressible quantum fluids beyond the conventional symmetry-breaking framework~\cite{wen_topological_1990,levin_stringnet_2005}. In particular, the Pfaffian state introduced by Moore and Read provides a minimal description of non-Abelian quasiparticles in a two-dimensional system~\cite{moore_nonabelions_1991} and has been widely discussed as a candidate for the even-denominator state at $\nu=5/2$~\cite{willett_observation_1987,greiter_paired_1992}. Microscopically, the Pfaffian implements chiral p-wave pairing at the level of the many-body wave function~\cite{greiter_paired_1991}, with particles forming correlated pairs that yield a gapped phase with non-Abelian topological order. Quasihole excitations carry localized Majorana zero modes~\cite{read_paired_2000}, leading to a ground-state degeneracy whose exchange statistics is non-commutative rather than simple phase factors. This non-Abelian structure enables quantum information to be encoded nonlocally in collective degrees of freedom~\cite{kitaev_faulttolerant_2003,nayak_nonabelian_2008}.

The prospect of fault-tolerant quantum computation based on non-Abelian FQH states has motivated sustained efforts to realize and control such phases in a variety of experimental settings. In electronic systems, the $\nu=5/2$ and other even-denominator states have been the primary focus, with tunneling~\cite{dolev_observation_2008,radu_quasiparticle_2008}, interferometric~\cite{willett_magneticfieldtuned_2013,willett_interference_2023,kim_aharonov_2026}, and thermal transport~\cite{banerjee_observation_2018} measurements reporting signatures consistent with non-Abelian topological order, although control and microscopic access remain challenging. In parallel, engineered quantum systems, such as cold-atom platforms, have emerged as highly controllable settings for exploring FQH physics, with several experiments recently realizing two-particle Laughlin states~\cite{clark_observation_2020,leonard_realization_2023,wang_realization_2024,lunt_realization_2024}; yet extending these approaches to larger systems and to phases with non-Abelian order requires operating in narrow parameter regimes at temperatures below small many-body gaps.

In this work, we leverage the single-atom control and full spatial resolution of our quantum gas microscope~\cite{bakr_quantum_2009} to directly engineer and probe a three-particle bosonic Pfaffian state using ultracold $^{87}\mathrm{Rb}$ atoms in an optical lattice. We identify a regime of the interacting Harper–Hofstadter Hamiltonian~\cite{harper_singleband_1955,hofstadter_energylevels_1976} on a $5\times 5$ lattice whose many-body ground state exhibits a 93.9\% overlap with the Pfaffian wave function and realize this state from the ground up. Synthetic magnetic fields are generated using Floquet engineering, enabling an adiabatic connection from a trivial initial state to the Pfaffian state via a Bayesian-optimized ramp. With site-resolved detection, we directly observe the pairing correlations emblematic of the Pfaffian state and further perform controlled drift measurements to probe its transport properties.

\section*{Three-particle Pfaffian state}
To place our experimental realization in context, we briefly outline the structure of the Pfaffian wave function underlying the state we prepare. In the continuum, the bosonic Pfaffian state can be written as a Laughlin Jastrow factor multiplied by a BCS pairing amplitude and projected onto fixed particle number~\cite{greiter_paired_1991},
\begin{align}
\label{eq:pfaffian_state}
    \Psi[z]&=\mathrm{Pf}\left(\frac{1}{z_i - z_j} \right)
    \prod_{i<j}^N (z_i - z_j)\, \prod_{i}^N \text{e}^{-\frac{1}{4 \ell_{\!B}^2}|z_i|^2},
\end{align}
where $z_i=x_i+\mathrm{i}y_i$ denote the particle coordinates in the complex plane and $\ell_{\!B}$ is the magnetic length. The Pfaffian ($\mathrm{Pf}$) is the antisymmetrized product of all possible pairings. It distinguishes the state from conventional Abelian quantum Hall liquids and underpins its non-Abelian properties.

\begin{figure}[t!!]
\centering
\includegraphics[width=\linewidth]{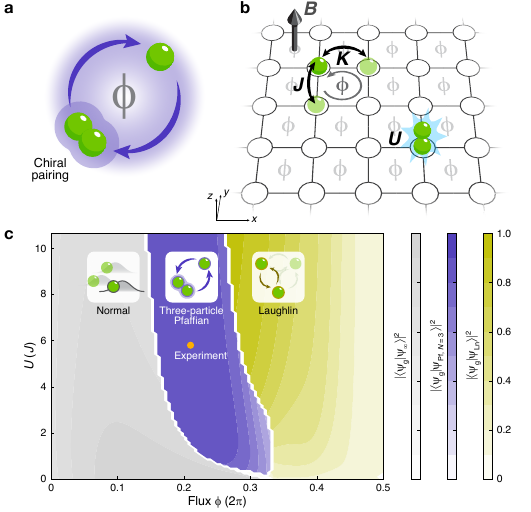}
\caption{\textbf{Three-particle Pfaffian state of ultracold bosons in an optical lattice.} 
\textbf{a}, The analytic wave function proposed in equation~\eqref{eq:our_state_N=3} describes a three-particle bosonic Pfaffian state. While the full Pfaffian wave function describes a p-wave paired many-body state, in the three-particle case one particle remains unpaired and the wave function is modified by a finite-size correction. The resulting wave function describes vortex motion between the pair and the unpaired single particle.
\textbf{b}, By engineering a synthetic magnetic field, we realize this state with three charge-neutral $^{87}\textrm{Rb}$ atoms in an optical lattice, confined to $5\times 5$ sites. The system is described by the interacting bosonic Harper–Hofstadter Hamiltonian (Methods), with effective magnetic flux per plaquette $\phi$, nearest-neighbor tunnelings $J$ along $y$ and $K$ along $x$, and on-site interaction $U$. 
\textbf{c}, Overlap of the ground state in the Harper–Hofstadter Hamiltonian with three analytic wave functions: the normal (topologically trivial) state (gray), the three-particle Pfaffian state (purple), and the $1/2$–Laughlin state (yellow) as a function of flux $\phi$ and interaction $U$. At the experimental parameters (orange dot) the ground state has a 93.9\% overlap with the three-particle Pfaffian state in equation~\eqref{eq:our_state_N=3} (Extended Data Fig.~\ref{ext_data:analytic_overlaps}).
}
\label{fig:Intro}
\end{figure}

For three particles, the Pfaffian pairs two particles while leaving the third unpaired (Fig.~\ref{fig:Intro}a). To connect the continuum construction in equation~\eqref{eq:pfaffian_state} to the finite lattice system in our experiment, we devise the three-particle Pfaffian wave function,
\begin{align}
\label{eq:our_state_N=3}
    \Psi[z]&=\mathcal{A}\left(\frac{z_1+z_2-2z_3}{z_1 - z_2} \right)
    \prod_{i<j}^3 (z_i - z_j)\, \prod_{i}^3 \text{e}^{-\frac{1}{4 \ell_{\!B}^2}|z_i|^2},
\end{align}
where $\mathcal{A}$ denotes antisymmetrization. The numerator in equation~\eqref{eq:our_state_N=3} describes a finite size correction, which compensates the reduction of angular momentum due to the pairing, as elaborated in Methods Section~\ref{sec:methods_analytic}.
Although this wave function describes a circular droplet in a continuous Landau level, it exhibits a 93.9\% overlap with the finite-size ground state of the Harper–Hofstadter Hamiltonian on the $5\times 5$ lattice targeted in our experiment (Fig.~\ref{fig:Intro}c). This large overlap reflects the pairing correlations shared by the two states.

\begin{figure*}[t!!]
\centering
\includegraphics[width=\textwidth]{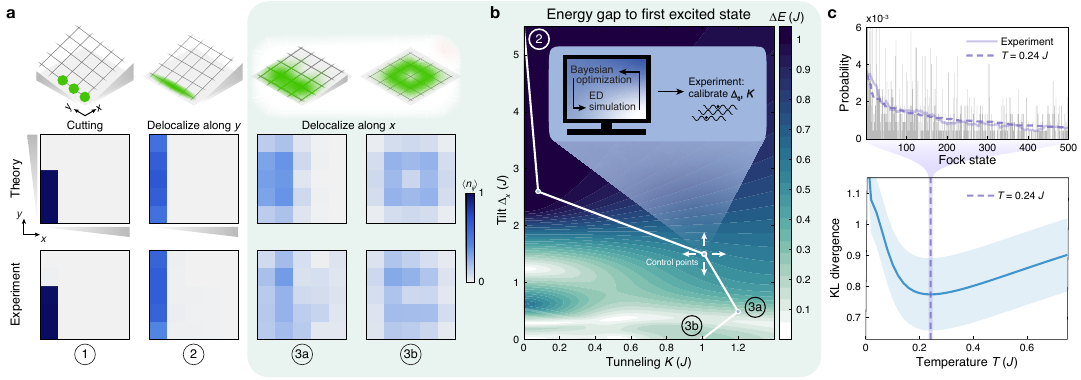}
\caption{
\textbf{Preparation of a minimal bosonic Pfaffian state.} 
\textbf{a}, Experimental sequence. Shown are schematics of the state in the Harper–Hofstadter Hamiltonian (top), the corresponding ground-state density from exact diagonalization (middle), and the measured in-situ density (bottom). The sequence consists of three stages: initialization of three localized atoms on a $5\times 5$ lattice, delocalization along $y$ into a 1D tube, and delocalization along 
$x$ into the final 2D Pfaffian state. Panels (3a) and (3b) show densities near the end of the final ramp stage.
\textbf{b}, Energy gap between the ground and first excited states as a function of the tunneling amplitude along $x$, $K$, and the residual tilt along $x$, $\Delta_x$, calculated by exact diagonalization. The experimental ramp trajectory is overlaid in white. The ramp consists of four segments whose endpoints are first determined by Bayesian optimization and then refined experimentally (Methods).
\textbf{c}, Temperature estimate from the Kullback–Leibler (KL) divergence between experiment and theory, yielding $T=0.24(1)~J$. The dark blue curve shows the KL divergence for the full dataset, and the shaded region indicates the standard deviation from bootstrap resampling. Top, subset of the experimental Fock-state distribution ordered by numerical probability (gray), compared with theory at the fitted temperature. The solid purple curve shows the 50-state running average of the experimental data, and the dashed purple curve the theory without averaging.}
\label{fig:Expsequence}
\end{figure*}

\section*{Preparation of the Pfaffian state}
To realize this state experimentally, we engineer a lattice Hamiltonian that captures the essential physics of FQH states. In electronic systems, FQH states including the Pfaffian arise from the interplay between strong interactions and the orbital motion induced by a magnetic field. In our neutral-atom quantum simulator, interactions arise naturally from the on-site repulsion $U$, while the effects of a magnetic field are implemented using Floquet engineering. The resulting dynamics are described by the Harper–Hofstadter Hamiltonian for interacting bosons on a lattice (Fig.~\ref{fig:Intro}b).

In the Harper–Hofstadter Hamiltonian, the magnetic field is encoded by a Peierls phase $\phi$ acquired during tunneling between lattice sites. To realize this phase, we engineer a synthetic magnetic field by applying a magnetic-field gradient and an optical running-wave lattice to drive Raman-assisted tunneling along the $x$ direction~\cite{tai_microscopy_2017}. This process imprints a spatially dependent complex tunneling phase $\phi$, analogous to the Aharonov–Bohm phase accumulated by a charged particle moving in a magnetic field, generating a synthetic magnetic field.

Our system provides independent control over the key Hamiltonian parameters required for adiabatic state preparation. The flux $\phi$ is set by the angle of the running-wave beams~\cite{tai_microscopy_2017}. Along $y$, a magnetic-field gradient controls the tilt $\Delta_y$, while the lattice depth sets the tunneling amplitude $J$. Along $x$, the detuning of the Raman drive controls the residual tilt $\Delta_x$, and the Raman drive power sets the tunneling amplitude $K$ (Methods). For most of the experiment the tunneling along $y$ is held fixed at $J/h = 33(1)$~Hz, where $h$ is Planck’s constant. This defines the natural energy scale of the system and the tunneling time $\tau = h/2\pi J = 4.8$~ms. The on-site interaction energy remains constant at $U = 5.76(2)~J$.

Leveraging our experimental flexibility, we take a bottom-up approach to quantum state engineering, where the system is coherently transferred from an easily prepared ground state to the target state by ramping the Hamiltonian parameters. In practice, however, engineering topologically protected states like the Pfaffian using adiabatic ramps presents a set of particular challenges. Adiabatically connecting a trivial state to a topologically ordered state requires traversing small excitation gaps, which in turn requires long ramp times to maintain adiabaticity. On the other hand, the prethermal nature of our driven system limits the coherence time, favoring shorter ramp times to reduce heating by the drive. Choosing parameters that balance these two competing restrictions is a major experimental challenge (Methods). 

Further complicating adiabatic state preparation is the sensitivity of the ramp to on-site disorder. The Raman beams that generate the synthetic magnetic field are also the main source of disorder, introducing potential variations comparable to the excitation gap at the most sensitive stage of the ramp ($\sim 0.1 - 0.2~J$; see Methods and Extended Data Fig.~\ref{ext_data:E_disorder}). This disorder can substantially modify the optimal ramp shape and, if not accounted for in the ramp design, increase the ramp duration required to maintain adiabaticity.

To address these challenges, we use Bayesian optimization~\cite{garnett_bayesian_2023} to design a ramp that optimally navigates the small excitation gaps and mitigates the negative effects of the Floquet drive. This method has recently been applied successfully to early stages of cold-atom experiments, including the rapid production of Bose–Einstein condensates~\cite{wigley_fast_2016,vendeiro_machinelearning_2022}.
Here we extend its use to the preparation of a topological many-body state, following the strategy outlined in~\cite{blatz_bayesian_2024}. Specifically, we apply it to the final stage of the sequence, where the excitation gap is smallest and the Floquet drive is active, by designing a simultaneous ramp of the tilt $\Delta_x$ and tunneling $K$ along $x$ that minimizes the time spent under driving.

The experimental sequence consists of three stages that connect an initially 1D system to the final 2D Pfaffian state (Fig.~\ref{fig:Expsequence}a). First, a digital micromirror device (DMD) isolates three adjacent atoms from the unity-filling shell of a Mott insulator. The DMD also projects confining walls that restrict the system to a $5\times 5$ lattice (Methods). At this stage, tunneling is suppressed along both $x$ and $y$, while a magnetic-field gradient along $y$ creates an energy offset of $2~J$ between neighboring sites. The atoms are then adiabatically delocalized along $y$ by ramping $J$ to its final value of $J/h=33(1)$~Hz and reducing $\Delta_y$ to form a 1D superfluid. In the final stage, we apply the Bayesian-optimized ramp to delocalize along $x$, simultaneously increasing $K$ from $0$ to $1~J$ and reducing $\Delta_x$ from $5.5~J$ to $0$. This stage is the most delicate, as the excitation gap narrows and tunneling along $x$ is driven by the Raman beams, making the ramp sensitive to both the reduced coherence time and the disorder introduced by the Floquet drive.

The optimized ramp is obtained by performing Bayesian optimization on a numerical simulation of the system to navigate the gap landscape shown in Fig.~\ref{fig:Expsequence}b. The ramp is parametrized into four linear segments of equal duration, giving the optimizer access to three control points.
Following the numerical optimization, we adapt the ramp to the experiment by experimentally calibrating the total ramp time and an overall scaling factor of the tilt $\Delta_x$ to maximize the probability of returning to the initial state after reversing the ramp (Methods). This two-step process results in the protocol indicated by the white line in Fig.~\ref{fig:Expsequence}b.

\begin{figure*}[t!!]
\centering
\includegraphics[width=\textwidth]{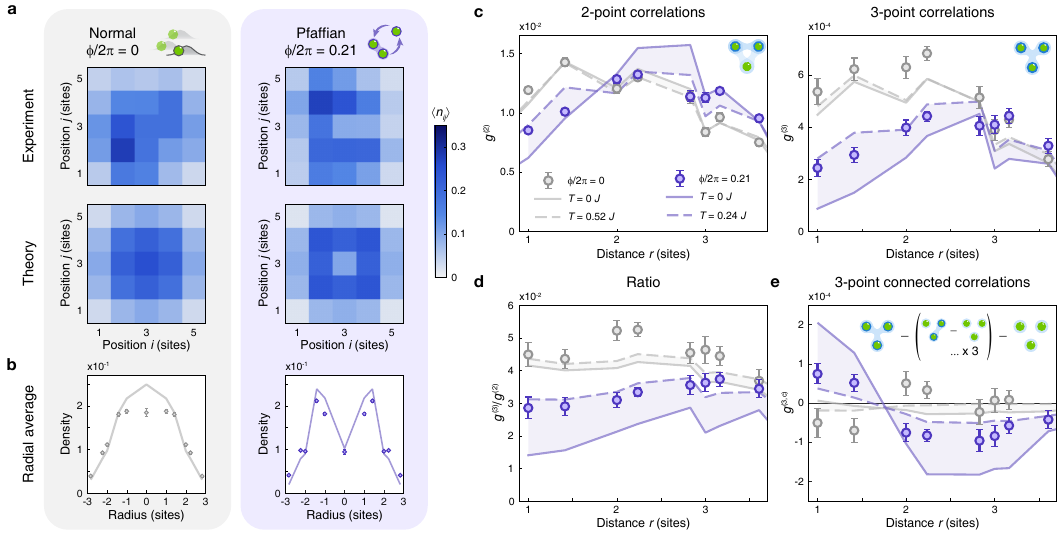}
\caption{
\textbf{Particle densities and correlations in the Pfaffian and normal regimes.} 
\textbf{a}, Measured density distributions on the $5 \times 5$ lattice (top) compared with ground-state theory (bottom). 
\textbf{b}, Radial average of the density distribution. For the three-particle Pfaffian state, defined in equation~\eqref{eq:our_state_N=3}, depletion at the central site is clearly visible, consistent with the chiral structure induced by the synthetic magnetic field. Solid lines show ground-state theory. Error bars denote s.e.m.
\textbf{c}, Two- and three-point density correlation functions probing the pairing structure of the Pfaffian state.
The data show clear suppression of $g^{(3)}$ at short distances, consistent with Pfaffian pairing, and agree well with theory (lines) at the estimated temperatures.
\textbf{d}, Ratio $g^{(3)} / g^{(2)}$, which quantifies the suppression of three-body correlations relative to two-body correlations.
\textbf{e}, Connected correlations, subtracting all lower-order contributions (equation~\eqref{eq:g3_fully_connected}).
This most sensitive measure demonstrates the genuine three-body correlated nature of the Pfaffian. Error bars in c–e denote bootstrap-estimated 68\% confidence intervals.
}
\label{fig:densities_and_correlations}
\end{figure*}

After completing the ramp, we project the state onto the number basis by quenching tunneling along both lattice directions and perform fluorescence imaging. Because imaging parity-projects multiply occupied sites~\cite{bakr_quantum_2009}, we post-select images containing three atoms on distinct lattice sites (Methods). Repeating the sequence builds up a probability distribution over number-basis configurations for the prepared state in the Pfaffian regime. For comparison, we also collect data in the normal regime at zero flux, using the Bayesian-optimized Pfaffian ramp as a starting point and applying the same experimental optimization procedure.

To quantify how closely the prepared state resembles the Pfaffian ground state, we extract a temperature by comparing the measured probability distribution with simulated thermal ensembles. The temperature is obtained by minimizing the Kullback–Leibler (KL) divergence between the experimental distribution and a Boltzmann-weighted distribution of eigenstates (Methods), yielding $T = 0.24(1)~J$ for the Pfaffian dataset and $T = 0.52(2)~J$ for the normal dataset (Fig.~\ref{fig:Expsequence}c).

To avoid introducing model-dependent assumptions about disorder, we compare the experimental distributions with simulations performed in the absence of disorder. This choice leads to a conservative temperature estimate, as disorder broadens the distribution in a way that mimics higher temperature (Methods). The extracted temperature nevertheless remains low enough to observe key Pfaffian signatures, which simulations indicate persist up to $T \approx 0.3~J$ (Extended Data Fig.~\ref{ext_data:connected_correlators}), while the normal-state behavior survives over a much broader temperature range owing to its larger excitation gap. This difference in robustness makes comparison of the two datasets meaningful despite their different temperatures.

\section*{Observation of Pfaffian pairing correlations}
The three-particle Pfaffian state is characterized by two particles forming a pair, leaving a third unpaired particle, shown in Fig.~\ref{fig:Intro}a and made explicit by the wave function in equation~\eqref{eq:our_state_N=3}. Crucially, this pairing is not driven by attractive interactions, but by a pairwise cancellation of repulsive terms introduced by the Pfaffian matrix. We therefore do not expect 
particle bunching at short distances, making the pairing challenging to probe. To uncover it, we characterize the prepared state using a sequence of increasingly selective observables, from qualitative signatures to direct probes of the pairing physics.

The measured density distribution already reveals a qualitative signature of the system's ground state in the presence of a synthetic magnetic field.
As shown in Fig.~\ref{fig:densities_and_correlations}a, introducing a flux $\phi/2\pi=0.21$ produces a dip in the center of the system compared to the normal state at zero flux.
The surrounding ring-shaped high-density region, visible in the radial average in Fig.~\ref{fig:densities_and_correlations}b, provides a first indication of the state's chiral structure.

To directly probe the pairing, we use the two- and three-point density correlation functions $g^{(2)}_{\mathbf{i}, \mathbf{j}} = \langle \hat{n}_\mathbf{i} \hat{n}_\mathbf{j} \rangle$, and $g^{(3)}_{\mathbf{i}, \mathbf{i'}, \mathbf{j}} = \langle \hat{n}_\mathbf{i} \hat{n}_\mathbf{i'} \hat{n}_\mathbf{j} \rangle$. Because the interactions are repulsive ($U>0$), $g^{(2)}$ is suppressed at very short distances in both the Pfaffian and normal states and therefore does not distinguish the paired structure on its own. The signature of pairing instead appears in the higher-order correlator $g^{(3)}$, which is suppressed at short distances in the Pfaffian state~\cite{Palm2021-BosonicPfaffianState}, despite the usual intuition that pairing enhances short-range correlations. Whereas the $\phi/2\pi=0$ normal state is governed only by contact repulsion, the Pfaffian state introduces an additional repulsion between the pair and the third particle, reducing the probability of finding all three particles close together.

To access these observables experimentally, we evaluate them on the parity-projected snapshots, for which $n_\mathbf{i} \in \{0, 1 \}$. Because doubly occupied sites cannot be distinguished from empty ones, we define a pair as two particles on nearest-neighbor sites, $|\mathbf{i} - \mathbf{i'}| = 1$, rather than on the same site. Fig.~\ref{fig:densities_and_correlations}c shows the correlators as a function of the distance $r = |\mathbf{i} - \mathbf{j}|$, with the three-point correlator averaged over all nearest neighbors $\mathbf{i'}$ of $\mathbf{i}$.

Comparison with the zero-flux normal state highlights the pairing signature in the Pfaffian state at $\phi/2\pi = 0.21$. The two-point correlations are similar in the two states, showing only weak suppression at the two shortest distances, whereas the three-point correlations are strongly suppressed in the Pfaffian state. This contrast is captured by the ratio $g^{(3)} / g^{(2)}$, shown in Fig.~\ref{fig:densities_and_correlations}d, which is strongly suppressed at short distances for the Pfaffian state and remains relatively flat for the normal state.

To isolate the genuine three-body contribution, we use the fully connected correlator
\begin{equation}
\begin{aligned}
    g^{(3, \: c)}_{\mathbf{i}, \mathbf{i'}, \mathbf{j}} &= \langle n_\mathbf{i} n_\mathbf{i'} n_\mathbf{j} \rangle - \frac{N-2}{N} \left( \langle n_\mathbf{i} n_\mathbf{i'} \rangle \langle n_\mathbf{j} \rangle + \mathrm{cyclic \ perm.} \right) \\
    & \quad \ + \: 2 \: \frac{(N-1)(N-2)}{N^2} \langle n_\mathbf{i} \rangle \langle n_\mathbf{i'} \rangle \langle n_\mathbf{j} \rangle,
    \label{eq:g3_fully_connected}
\end{aligned}
\end{equation}
which subtracts all lower-order contributions from $g^{(3)}_{\mathbf{i}, \mathbf{i'}, \mathbf{j}} = \langle n_\mathbf{i} n_\mathbf{i'} n_\mathbf{j} \rangle$. As is evident from the analytic wave functions in equations~(\ref{eq:Normal}-\ref{eq:Laughlin}), connected three-body terms are intrinsic to Pfaffian pairing, whereas the normal state contains only independent-particle contributions and the Laughlin state only two-body correlations. Consistent with this expectation, we observe finite connected three-body correlations in the Pfaffian regime, while they remain near zero in the normal state at $\phi/2\pi=0$ (Fig.~\ref{fig:densities_and_correlations}e). Taken together, the measured densities and correlation functions provide direct evidence for the pairing structure of the Pfaffian state and agree with the ground-state properties of the Harper–Hofstadter Hamiltonian.

\section*{Analysis of particle configurations}
The site-resolved imaging of our quantum gas microscope provides direct access to the spatial structure of the prepared state, allowing us to probe the paired nature of the Pfaffian wave function in detail. Complementing the correlations in Fig.~\ref{fig:densities_and_correlations}, which average over configurations with identical relative separations, we now use the exact particle positions to distinguish behavior in the central $3\times3$ region (bulk) from that near the edges.

\begin{figure}[h!]
\includegraphics[width=\columnwidth]{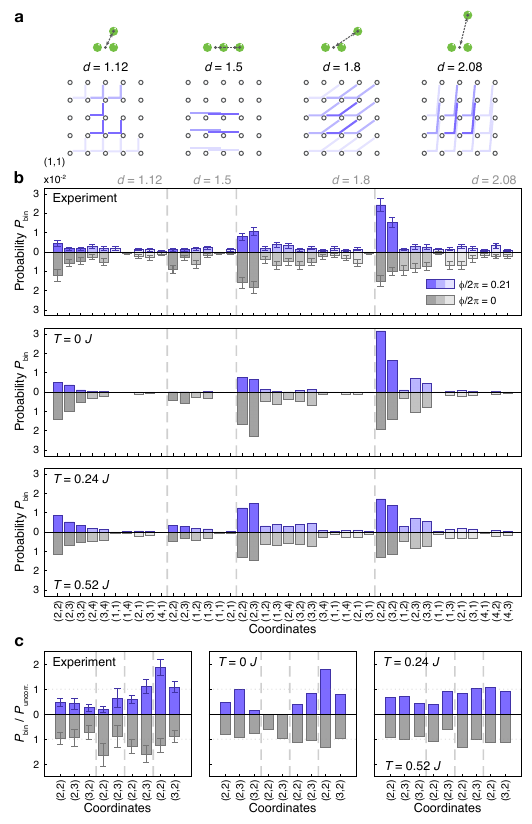}
\caption{
\textbf{Analysis of binned particle configurations.} Site-resolved imaging probes the paired structure of the Pfaffian wave function using the exact particle positions.
\textbf{a}, Images containing a nearest-neighbor pair are grouped into bins according to the bond-center distance $d$, defined as the distance between the third particle and the pair’s center of mass. Each bin is represented by a line connecting three lattice sites on the $5\times5$ grid. Configurations related by mirror or rotational symmetry are assigned to the same bin. Colors indicate the number of particles in the bulk: three (dark purple), two (mid purple), or one or zero (light purple). Bins are labeled by the ($x,y$) coordinate of the left-most site in the representative configuration.
\textbf{b}, Probability of each bin for the Pfaffian ($\phi/2\pi=0.21$) and normal ($\phi/2\pi=0$) states. At short bond-center distances, bulk bins are suppressed for the Pfaffian relative to the normal state, consistent with pairing. Bins containing at least one edge particle remain unlikely at all distances.
\textbf{c}, Bin probabilities normalized by those of an uncorrelated distribution reproducing the measured density. Only bulk bins are shown, with gray dashed lines separating bins of increasing bond-center distance. The Pfaffian state exhibits strong suppression at short bond-center distances, whereas the normal state remains comparatively uniform. Error bars denote bootstrap-estimated 68\% confidence intervals.
}
\label{fig:binning}
\end{figure}

We focus on configurations containing a nearest-neighbor pair and characterize them by the distance between the pair’s center of mass and the third particle, which we define as the bond-center distance. Images are grouped into bins according to the arrangement of particles on the $5\times 5$ lattice. In Fig.~\ref{fig:binning}a, each line connecting three lattice sites denotes one bin, and configurations related by mirror or rotational symmetry are assigned to that same bin. Restricting the analysis to the four shortest bond-center distances captures the relevant pairing physics (Fig.~\ref{fig:binning}a, top row).

The binned probabilities reveal a clear distinction between the Pfaffian and normal states (Fig.~\ref{fig:binning}b). For the Pfaffian state ($\phi/2\pi=0.21$), configurations with all particles in the bulk (dark purple) are strongly suppressed at short bond-center distances and increase as the third particle moves away from the nearest-neighbor pair. This behavior indicates that the third particle tends to remain a finite distance from the pair, consistent with the pairing structure of the Pfaffian wave function. Configurations involving edge sites remain unlikely across all distances, reflecting the tendency for particles to avoid the boundaries. In contrast, the normal state ($\phi/2\pi=0$) exhibits enhanced probability for short bond-center distances in the bulk, consistent with the absence of pairing correlations.

Comparison with theory suggests that, for the Pfaffian data, bulk configurations remain close to the ground state while excitations appear predominantly at the edge. In the bulk, the data are most consistent with ground-state predictions, showing strong suppression at short bond-center distances and a sharp increase at the largest distance, whereas configurations involving edge sites agree better with finite-temperature theory ($T=0.24~J$). This indicates that pairing correlations dominate in the bulk, while excitations, modeled as finite temperature, appear primarily through the occupation of edge sites. Disorder, which is not included in the simulations, may also contribute to this behavior.

The paired nature of the state becomes especially clear in Fig.~\ref{fig:binning}c, where bin probabilities are normalized by those of an uncorrelated distribution reproducing the measured density. Restricting to bulk configurations, the Pfaffian state shows a pronounced suppression at short bond-center distances, whereas the normal state remains comparatively uniform. These observations provide direct spatial evidence of the pairing correlations that define the Pfaffian state.

\begin{figure*}[t!!]
\centering
\includegraphics[width=\textwidth]{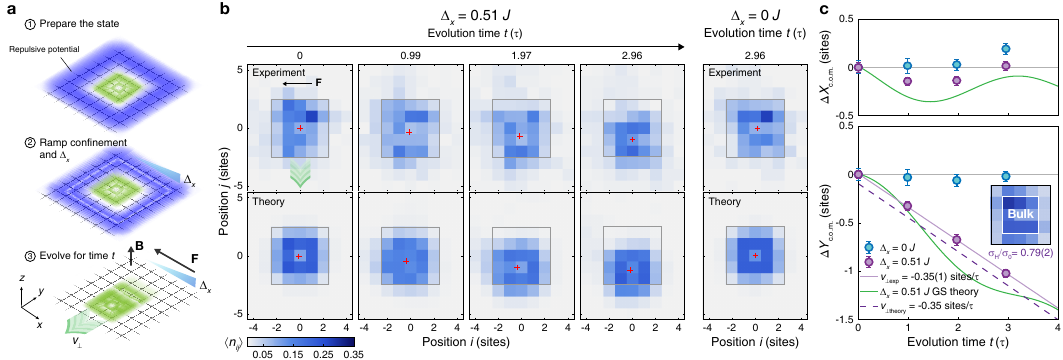}
\caption{
\textbf{The Hall drift experiment.} Transport response of the Pfaffian state probed through center-of-mass drift.
\textbf{a}, Experimental protocol: (1) prepare the Pfaffian state within the repulsive walls; (2) lower the confinement and apply a force along the $x$ direction; (3) measure the density profile after an evolution time $t$.
\textbf{b}, Measured density profiles with an applied tilt (top) compared with theory (bottom). In the presence of a force, the density drifts in the transverse $y$ direction, whereas it remains stationary over the same evolution time when no force is applied (right). 
\textbf{c}, Transverse center-of-mass drift. The drift velocity $v_\perp$ is extracted from a linear fit to the transverse center-of-mass drift $\Delta Y_{\mathrm{c.o.m.}}(t)$ after release (purple solid line). It is in good quantitative agreement with the theoretical drift velocity (purple dashed line) extracted by linear fit to the center-of-mass of the ground state at long evolution times (solid green line). The bulk density is obtained from the central $3\times3$ region of the in-situ density, yielding a Hall conductivity $\sigma_\mathrm{H}/\sigma_0 = 0.79(2)$. Error bars denote s.e.m. 
}
\label{fig:hall_drift}
\end{figure*}

\section*{Probing the transport response of the Pfaffian state}
Finally, we probe the transport response of the Pfaffian state through a Hall drift measurement, complementing the in-situ observables discussed above~\cite{leblanc_observation_2012,aidelsburger_measuring_2014}. Theoretical work has shown that few-particle FQH states can exhibit quantized Hall conductivity despite their small system sizes~\cite{motruk_detecting_2020,repellin_fractional_2020}. We measure the Hall response by monitoring the transverse center-of-mass drift under a weak force. The state is first prepared using the same protocol as for the in-situ measurements. We then lower the confinement walls and ramp on a tilt along the $x$ direction to $\Delta_x = 0.51(1)~J$ in $1~\tau$ (Fig.~\ref{fig:hall_drift}a). In response, the state acquires a transverse drift velocity $v_\perp$ characteristic of the Hall effect.

We measure density profiles at different evolution times $t$ and use the resulting center-of-mass motion to extract the Hall conductivity (Fig.~\ref{fig:hall_drift}b). In the presence of a static force along $x$, the ring-shaped density profile shifts by approximately one lattice site in the transverse $y$ direction, whereas it remains stationary when no force is applied. The measured density profiles are in good qualitative agreement with theory.

We determine the transverse drift velocity $v_\perp$ from the slope of the $y$ center-of-mass displacement. Combining this with the bulk density $\rho_{\mathrm{bulk}}$ obtained from the in-situ measurements, we extract a Hall conductivity $\sigma_\mathrm{H} / \sigma_0 = 2\pi \rho_{\mathrm{bulk}} v_\perp / \Delta_x = 0.79(2)$, where $\sigma_0$ is the conductivity quantum. This is lower than the finite-size expectation $\sigma_\mathrm{H} / \sigma_0 = 0.94$ for the $\nu=1$ Pfaffian state. We attribute the discrepancy primarily to imperfect state preparation and disorder. Excitations generated during the preparation reduce the bulk density, while the disorder landscape during the drift modifies the local transverse response (Methods). These results establish center-of-mass drift as a practical probe of Hall conductivity in few-atom systems and enable more quantitative characterization of FQH states in optical lattices.

\section*{Conclusion and outlook}
The versatility and particle-resolved control of cold-atom quantum simulators make them an appealing platform for realizing strongly correlated topological states, enabling direct probes of microscopic quantities and, perhaps eventually, new approaches to fault-tolerant quantum computation~\cite{goldman_topological_2016}. Although minimal Laughlin states have been realized in related systems, the bottom-up preparation of FQH states with non-Abelian topological order, such as the Pfaffian state, has remained an outstanding experimental challenge. By directly observing the characteristic suppression of $g^{(3)}$ correlations in the finite-size regime expected for the Pfaffian state, close to filling factor $\nu = 1$, we provide strong evidence for the realization of a topological state linked to non-Abelian anyonic statistics. The measurement of the transport response provides further evidence for the robustness of the engineered ground state and a basis for future measurements of the Hall conductivity~\cite{repellin_fractional_2020}. Our work establishes bottom-up engineering as a route to probing topological observables that have so far remained experimentally inaccessible.

There are still challenges to overcome before such measurements can be made in our system. We identify potential disorder as the main limitation of the current approach, an insight that has guided the strategies developed here to mitigate its effects and simplify future experiments. Although disorder is a common challenge in cold-atom platforms, here the challenge is compounded by its interplay with the Floquet drive, which imposes conflicting constraints on the preparation ramp: slower ramps reduce diabatic errors, whereas faster ramps are needed to limit Floquet heating. In this work, we address this trade-off using Bayesian optimization to design a simultaneous ramp of tilt and tunneling that is robust to disorder while remaining sufficiently fast to limit Floquet heating. This approach can be extended to larger systems and further improved using diabatic ramp design~\cite{blatz_bayesian_2024,wu_optimal_2025}. The methods described in Methods Section~\ref{sec:EandUDisorder} could also be used to map the disorder landscape and tailor ramps to the measured potential. Future experimental upgrades could replace the disordered Raman beams with a programmable spatial light modulator with high frame rate and low disorder~\cite{wei_10megahertz_2026}. Reducing the disorder would enable faster and more robust state preparation, as well as larger system sizes.

While measuring topological observables in such a small system would be a testament to the robustness of topological order, scaling beyond the minimal system size, either directly or by adiabatically connecting many smaller systems~\cite{palm_growing_2024}, would broaden the scope of the results and enable additional types of measurements. A next major goal is to measure topological observables, including plateaus in the Hall conductivity~\cite{repellin_fractional_2020}, the edge excitation spectrum~\cite{zache_entanglement_2022,binanti_spectroscopy_2024,nardin_quantumnonlinear_2024}, and fractionally charged excitations~\cite{raciunas_creating_2018,umucalilar_bulk_2023,palm_growing_2024}. Looking further ahead, an important prospect is the direct control of non-Abelian anyonic excitations~\cite{kapit_nonabelian_2012,wang_measurable_2022,palm_interferometric_2025}. Progress towards this direction would mark an important step towards harnessing Pfaffian states for quantum computation.

\subsection*{Acknowledgments}
We thank I.~Carusotto, A.~Douglas, N.~Goldman, H.-Y.~Hu, A.~M.~Kaufman, C.~Kokail, M.~Lebrat, Y.~K.~Lee, J.~L\'eonard, M.~Lewenstein, A.~Lukin, A.~Nardin, F.~A.~Palm, C.~Repellin, N.~Suri, M.~E.~Tai, and R.~O.~Umucal{\i}lar for insightful discussions. 

\subsection*{Funding}
This work was supported by the National Science Foundation under Grant Number 2317134, the Army Research Office Grant Number W911NF-20-1-0163, the U.S. DoE Office of Science award No. DE-AC02-05CH11231, and the Gordon and Betty Moore Foundation through Grant GMBF 11521. A.~B., T.~B., F.~G. acknowledge funding by the Deutsche Forschungsgemeinschaft (DFG, German Research Foundation) under Germany's Excellence Strategy -- EXC-2111 -- 390814868. M.~Greiter was supported by the Deutsche Forschungsgemeinschaft (DFG, German Research Foundation) through the Würzburg--Dresden Cluster of Excellence \textit{ctd.qmat} -- Complexity and Topology and Dynamics in Quantum Matter (EXC 2147, project-id 390858490).

\subsection*{Author contributions}
J.~K., P.~S., Y.~L., and A.~Z. performed the experiment and collected data. J.~K., P.~S., Y.~L., T.~B., and A.~Z analyzed data. J.~K., P.~S., Y.~L., T.~B., A.~Z, and B.~B.-H. performed numerical simulations. T.~B., M.~Greiter, F.~G., and A.~B. performed theoretical analysis. M.~Greiner, F.~G., M.~Greiter, and A.~B. supervised the project. All authors contributed to the interpretation of the results and writing of the manuscript.

\subsection*{Competing interests}
M.~Greiner is a co-founder, shareholder, and consultant of QuEra Computing. All other authors declare no competing interests.
\\
\\
$^{*}$~These authors contributed equally to this work; \\
$^{\dagger}$~Present address: JILA and Department of Physics, University of Colorado, Boulder, Colorado 80309, USA; \\
$^{\ddagger}$~Present address: Laboratoire Kastler Brossel, Coll\`ege de France, CNRS, ENS-Universit\'e PSL, Sorbonne Universit\'e, 75005 Paris, France; \\
$^{\S}$~Corresponding author: mgreiner@g.harvard.edu

\bibliography{references}

@PREAMBLE{
 "\providecommand{\noopsort}[1]{}" 
 # "\providecommand{\singleletter}[1]{#1}%" 
}

@article{aidelsburger_measuring_2014,
   title={Measuring the {Chern} number of {Hofstadter} bands with ultracold bosonic atoms},
   volume={11},
   ISSN={1745-2481},
   url={http://dx.doi.org/10.1038/nphys3171},
   DOI={10.1038/nphys3171},
   number={2},
   journal={Nature Physics},
   publisher={Springer Science and Business Media LLC},
   author={Aidelsburger, M. and Lohse, M. and Schweizer, C. and Atala, M. and Barreiro, J. T. and Nascimbène, S. and Cooper, N. R. and Bloch, I. and Goldman, N.},
   year={2014},
   month=dec, pages={162–166} 
   }

@article{bakr_quantum_2009,
  title = {A Quantum Gas Microscope for Detecting Single Atoms in a {{Hubbard-regime}} Optical Lattice},
  author = {Bakr, Waseem S. and Gillen, Jonathon I. and Peng, Amy and F{\"o}lling, Simon and Greiner, Markus},
  year = 2009,
  month = nov,
  journal = {Nature},
  volume = {462},
  number = {7269},
  pages = {74--77},
  publisher = {Nature Publishing Group},
  issn = {1476-4687},
  doi = {10.1038/nature08482},
  urldate = {2023-12-18},
  copyright = {2009 Macmillan Publishers Limited. All rights reserved},
  langid = {english}
}

@article{banerjee_observation_2018,
  title = {Observation of Half-Integer Thermal {{Hall}} Conductance},
  author = {Banerjee, Mitali and Heiblum, Moty and Umansky, Vladimir and Feldman, Dima E. and Oreg, Yuval and Stern, Ady},
  year = 2018,
  month = jul,
  journal = {Nature},
  volume = {559},
  number = {7713},
  pages = {205--210},
  publisher = {Nature Publishing Group},
  issn = {1476-4687},
  doi = {10.1038/s41586-018-0184-1},
  urldate = {2026-02-18},
  copyright = {2018 Macmillan Publishers Ltd., part of Springer Nature},
  langid = {english}
}

@article{binanti_spectroscopy_2024,
  title = {Spectroscopy of edge and bulk collective modes in fractional {Chern} insulators},
  author = {Binanti, F. and Goldman, N. and Repellin, C.},
  journal = {Phys. Rev. Res.},
  volume = {6},
  issue = {1},
  pages = {L012054},
  numpages = {6},
  year = {2024},
  month = {Mar},
  publisher = {American Physical Society},
  doi = {10.1103/PhysRevResearch.6.L012054},
  url = {https://link.aps.org/doi/10.1103/PhysRevResearch.6.L012054}
}

@article{blatz_bayesian_2024,
  title = {Bayesian Optimization for Robust State Preparation in Quantum Many-Body Systems},
  author = {Blatz, Tizian and Kwan, Joyce and L{\'e}onard, Julian and Bohrdt, Annabelle},
  year = 2024,
  month = jun,
  journal = {Quantum},
  volume = {8},
  pages = {1388},
  publisher = {Verein zur F\"orderung des Open Access Publizierens in den Quantenwissenschaften},
  doi = {10.22331/q-2024-06-27-1388},
  urldate = {2025-09-05},
  langid = {british}
}

@article{clark_observation_2020,
  title = {Observation of {{Laughlin}} States Made of Light},
  author = {Clark, Logan W. and Schine, Nathan and Baum, Claire and Jia, Ningyuan and Simon, Jonathan},
  year = 2020,
  month = jun,
  journal = {Nature},
  volume = {582},
  number = {7810},
  pages = {41--45},
  publisher = {Nature Publishing Group},
  issn = {1476-4687},
  doi = {10.1038/s41586-020-2318-5},
  urldate = {2026-02-18},
  copyright = {2020 The Author(s), under exclusive licence to Springer Nature Limited},
  langid = {english}
}

@article{dolev_observation_2008,
  title = {Observation of a Quarter of an Electron Charge at the {$\nu$} = 5/2 Quantum {{Hall}} State},
  author = {Dolev, M. and Heiblum, M. and Umansky, V. and Stern, Ady and Mahalu, D.},
  year = 2008,
  month = apr,
  journal = {Nature},
  volume = {452},
  number = {7189},
  pages = {829--834},
  publisher = {Nature Publishing Group},
  issn = {1476-4687},
  doi = {10.1038/nature06855},
  urldate = {2026-02-18},
  copyright = {2008 Springer Nature Limited},
  langid = {english}
}

@book{garnett_bayesian_2023,
  title = {Bayesian {Optimization}},
  author = {Garnett, Roman},
  year = 2023,
  publisher = {Cambridge University Press},
  address = {Cambridge},
  doi = {10.1017/9781108348973},
  urldate = {2023-10-11},
  isbn = {978-1-108-42578-0}
}

@article{goldman_topological_2016,
  title = {Topological quantum matter with ultracold gases in optical lattices},
  author = {Goldman, N. and Budich, J. and Zoller, P.},
  journal = {Nature Phys},
  volume = {12},
  pages = {639–645},
  numpages = {7},
  year = {2016},
  month = {Jun},
  publisher = {Springer Nature},
  doi = {https://doi.org/10.1038/nphys3803},
  url = {https://www.nature.com/articles/nphys3803}
}

@article{greiter_paired_1991,
  title = {Paired {{Hall}} State at Half Filling},
  author = {Greiter, Martin and Wen, Xiao-Gang and Wilczek, Frank},
  year = 1991,
  month = jun,
  journal = {Phys. Rev. Lett.},
  volume = {66},
  number = {24},
  pages = {3205--3208},
  publisher = {American Physical Society},
  doi = {10.1103/PhysRevLett.66.3205},
  urldate = {2026-02-18}
}

@article{greiter_paired_1992,
  title = {Paired {{Hall}} States},
  author = {Greiter, Martin and Wen, X. G. and Wilczek, Frank},
  year = 1992,
  month = may,
  journal = {Nuclear Physics B},
  volume = {374},
  number = {3},
  pages = {567--614},
  issn = {0550-3213},
  doi = {10.1016/0550-3213(92)90401-V},
  urldate = {2026-02-18}
}

@article{harper_singleband_1955,
doi = {10.1088/0370-1298/68/10/304},
url = {https://doi.org/10.1088/0370-1298/68/10/304},
year = {1955},
month = {oct},
publisher = {},
volume = {68},
number = {10},
pages = {874},
author = {P G Harper},
title = {Single Band Motion of Conduction Electrons in a Uniform Magnetic Field},
journal = {Proceedings of the Physical Society. Section A}
}

@article{hofstadter_energylevels_1976,
  title = {Energy levels and wave functions of {B}loch electrons in rational and irrational magnetic fields},
  author = {Hofstadter, Douglas R.},
  journal = {Phys. Rev. B},
  volume = {14},
  issue = {6},
  pages = {2239--2249},
  numpages = {0},
  year = {1976},
  month = {Sep},
  publisher = {American Physical Society},
  doi = {10.1103/PhysRevB.14.2239},
  url = {https://link.aps.org/doi/10.1103/PhysRevB.14.2239}
}

@article{hudomal_bosonic_2019,
  title = {Bosonic fractional quantum {Hall} states in driven optical lattices},
  author = {Hudomal, Ana and Regnault, Nicolas and Vasi\ifmmode \acute{c}\else \'{c}\fi{}, Ivana},
  journal = {Phys. Rev. A},
  volume = {100},
  issue = {5},
  pages = {053624},
  numpages = {9},
  year = {2019},
  month = {Nov},
  publisher = {American Physical Society},
  doi = {10.1103/PhysRevA.100.053624},
  url = {https://link.aps.org/doi/10.1103/PhysRevA.100.053624}
}

@article{kapit_nonabelian_2012,
  title = {Non-{A}belian Braiding of Lattice Bosons},
  author = {Kapit, Eliot and Ginsparg, Paul and Mueller, Erich},
  journal = {Phys. Rev. Lett.},
  volume = {108},
  issue = {6},
  pages = {066802},
  numpages = {5},
  year = {2012},
  month = {Feb},
  publisher = {American Physical Society},
  doi = {10.1103/PhysRevLett.108.066802},
  url = {https://link.aps.org/doi/10.1103/PhysRevLett.108.066802}
}

@article{kim_aharonov_2026,
  title = {Aharonov--{{Bohm}} Interference in Even-Denominator Fractional Quantum {{Hall}} States},
  author = {Kim, Jehyun and Dev, Himanshu and Shaer, Amit and Kumar, Ravi and Ilin, Alexey and Haug, Andr{\'e} and Iskoz, Shelly and Watanabe, Kenji and Taniguchi, Takashi and Mross, David F. and Stern, Ady and Ronen, Yuval},
  year = 2026,
  month = jan,
  journal = {Nature},
  volume = {649},
  number = {8096},
  pages = {323--329},
  publisher = {Nature Publishing Group},
  issn = {1476-4687},
  doi = {10.1038/s41586-025-09891-2},
  urldate = {2026-03-13},
  copyright = {2026 The Author(s)},
  langid = {english}
}

@article{kitaev_faulttolerant_2003,
  title = {Fault-Tolerant Quantum Computation by Anyons},
  author = {Kitaev, A. {\relax Yu}.},
  year = 2003,
  month = jan,
  journal = {Annals of Physics},
  volume = {303},
  number = {1},
  pages = {2--30},
  issn = {0003-4916},
  doi = {10.1016/S0003-4916(02)00018-0},
  urldate = {2026-03-10}
}

@article{laughlin_anomalous_1983,
  title = {Anomalous Quantum {Hall} Effect: {A}n Incompressible Quantum Fluid with Fractionally Charged Excitations},
  shorttitle = {Anomalous {{Quantum Hall Effect}}},
  author = {Laughlin, R. B.},
  year = 1983,
  month = may,
  journal = {Phys. Rev. Lett.},
  volume = {50},
  number = {18},
  pages = {1395--1398},
  publisher = {American Physical Society},
  doi = {10.1103/PhysRevLett.50.1395},
  urldate = {2026-02-01}
}

@article{leblanc_observation_2012,
   title={Observation of a superfluid {H}all effect},
   volume={109},
   ISSN={1091-6490},
   url={http://dx.doi.org/10.1073/pnas.1202579109},
   DOI={10.1073/pnas.1202579109},
   number={27},
   journal={Proceedings of the National Academy of Sciences},
   publisher={Proceedings of the National Academy of Sciences},
   author={LeBlanc, Lindsay J. and Jiménez-García, Karina and Williams, Ross A. and Beeler, Matthew C. and Perry, Abigail R. and Phillips, William D. and Spielman, Ian B.},
   year={2012},
   month=jun, pages={10811–10814} 
}

@article{leonard_realization_2023,
  title = {Realization of a Fractional Quantum {{Hall}} State with Ultracold Atoms},
  author = {L{\'e}onard, Julian and Kim, Sooshin and Kwan, Joyce and Segura, Perrin and Grusdt, Fabian and Repellin, C{\'e}cile and Goldman, Nathan and Greiner, Markus},
  year = 2023,
  month = jul,
  journal = {Nature},
  volume = {619},
  number = {7970},
  pages = {495--499},
  publisher = {Nature Publishing Group},
  issn = {1476-4687},
  doi = {10.1038/s41586-023-06122-4},
  urldate = {2026-02-18},
  copyright = {2023 The Author(s), under exclusive licence to Springer Nature Limited},
  langid = {english}
}

@article{levin_stringnet_2005,
  title = {String-Net Condensation: {{A}} Physical Mechanism for Topological Phases},
  shorttitle = {String-Net Condensation},
  author = {Levin, Michael A. and Wen, Xiao-Gang},
  year = 2005,
  month = jan,
  journal = {Phys. Rev. B},
  volume = {71},
  number = {4},
  pages = {045110},
  issn = {1098-0121, 1550-235X},
  doi = {10.1103/PhysRevB.71.045110},
  urldate = {2026-01-30},
  copyright = {http://link.aps.org/licenses/aps-default-license},
  langid = {english}
}

@article{lunt_realization_2024,
  title = {Realization of a {L}aughlin State of Two Rapidly Rotating Fermions},
  author = {Lunt, Philipp and Hill, Paul and Reiter, Johannes and Preiss, Philipp M. and Ga{\l}ka, Maciej and Jochim, Selim},
  year = 2024,
  month = dec,
  journal = {Phys. Rev. Lett.},
  volume = {133},
  number = {25},
  pages = {253401},
  publisher = {American Physical Society},
  doi = {10.1103/PhysRevLett.133.253401},
  urldate = {2026-02-18}
}

@article{ma_photonassisted_2011,
  title = {Photon-Assisted Tunneling in a Biased Strongly Correlated {B}ose Gas},
  author = {Ma, Ruichao and Tai, M. Eric and Preiss, Philipp M. and Bakr, Waseem S. and Simon, Jonathan and Greiner, Markus},
  journal = {Phys. Rev. Lett.},
  volume = {107},
  issue = {9},
  pages = {095301},
  numpages = {4},
  year = {2011},
  month = {Aug},
  publisher = {American Physical Society},
  doi = {10.1103/PhysRevLett.107.095301},
  url = {https://link.aps.org/doi/10.1103/PhysRevLett.107.095301}
}

@article{moore_nonabelions_1991,
  title = {Nonabelions in the Fractional Quantum {H}all Effect},
  author = {Moore, Gregory and Read, Nicholas},
  year = 1991,
  month = aug,
  journal = {Nuclear Physics B},
  volume = {360},
  number = {2},
  pages = {362--396},
  issn = {0550-3213},
  doi = {10.1016/0550-3213(91)90407-O},
  urldate = {2026-02-01}
}

@article{motruk_detecting_2020,
  title = {Detecting Fractional {C}hern Insulators in Optical Lattices through Quantized Displacement},
  author = {Motruk, Johannes and Na, Ilyoun},
  journal = {Phys. Rev. Lett.},
  volume = {125},
  issue = {23},
  pages = {236401},
  numpages = {6},
  year = {2020},
  month = {Dec},
  publisher = {American Physical Society},
  doi = {10.1103/PhysRevLett.125.236401},
  url = {https://link.aps.org/doi/10.1103/PhysRevLett.125.236401}
}

@article{nardin_quantumnonlinear_2024,
  title = {Quantum Nonlinear Optics on the Edge of a Few-Particle Fractional Quantum {H}all Fluid in a Small Lattice},
  author = {Nardin, Alberto and De Bernardis, Daniele and Umucal\ifmmode \imath \else \i \fi{}lar, Rifat Onur and Mazza, Leonardo and Rizzi, Matteo and Carusotto, Iacopo},
  journal = {Phys. Rev. Lett.},
  volume = {133},
  issue = {18},
  pages = {183401},
  numpages = {8},
  year = {2024},
  month = {Oct},
  publisher = {American Physical Society},
  doi = {10.1103/PhysRevLett.133.183401},
  url = {https://link.aps.org/doi/10.1103/PhysRevLett.133.183401}
}

@article{nayak_nonabelian_2008,
  title = {Non-{{Abelian}} Anyons and Topological Quantum Computation},
  author = {Nayak, Chetan and Simon, Steven H. and Stern, Ady and Freedman, Michael and Das Sarma, Sankar},
  year = 2008,
  month = sep,
  journal = {Rev. Mod. Phys.},
  volume = {80},
  number = {3},
  pages = {1083--1159},
  publisher = {American Physical Society},
  doi = {10.1103/RevModPhys.80.1083},
  urldate = {2026-02-18}
}

@article{Palm2021-BosonicPfaffianState,
  title = {Bosonic {{Pfaffian}} State in the {{Hofstadter-Bose-Hubbard}} Model},
  author = {Palm, F. A. and Buser, M. and L{\'e}onard, J. and Aidelsburger, M. and Schollw{\"o}ck, U. and Grusdt, F.},
  year = {2021},
  month = apr,
  journal = {Physical Review B},
  volume = {103},
  number = {16},
  pages = {L161101},
  publisher = {American Physical Society},
  doi = {10.1103/PhysRevB.103.L161101},
  urldate = {2023-11-06},
}

@article{palm_growing_2024,
  title = {Growing extended {L}aughlin states in a quantum gas microscope: {A} patchwork construction},
  author = {Palm, F. A. and Kwan, J. and Bakkali-Hassani, B. and Greiner, M. and Schollw\"ock, U. and Goldman, N. and Grusdt, F.},
  journal = {Phys. Rev. Res.},
  volume = {6},
  issue = {1},
  pages = {013198},
  numpages = {17},
  year = {2024},
  month = {Feb},
  publisher = {American Physical Society},
  doi = {10.1103/PhysRevResearch.6.013198},
  url = {https://link.aps.org/doi/10.1103/PhysRevResearch.6.013198}
}

@misc{palm_interferometric_2025,
      title={Interferometric Braiding of Anyons in {C}hern Insulators}, 
      author={Felix A. Palm and Nader Mostaan and Nathan Goldman and Fabian Grusdt},
      year={2025},
      eprint={2511.09445},
      archivePrefix={arXiv},
      primaryClass={quant-ph},
      url={https://arxiv.org/abs/2511.09445}, 
}

@article{preiss_strongly_2015,
  title = {Strongly Correlated Quantum Walks in Optical Lattices},
  author = {Preiss, Philipp M. and Ma, Ruichao and Tai, M. Eric and Lukin, Alexander and Rispoli, Matthew and Zupancic, Philip and Lahini, Yoav and Islam, Rajibul and Greiner, Markus},
  year = 2015,
  month = mar,
  journal = {Science},
  volume = {347},
  number = {6227},
  pages = {1229--1233},
  publisher = {American Association for the Advancement of Science},
  doi = {10.1126/science.1260364},
  urldate = {2026-03-10}
}

@article{raciunas_creating_2018,
  title = {Creating, probing, and manipulating fractionally charged excitations of fractional {C}hern insulators in optical lattices},
  author = {Ra\ifmmode \check{c}\else \v{c}\fi{}i\ifmmode \bar{u}\else \={u}\fi{}nas, Mantas and \"Unal, F. Nur and Anisimovas, Egidijus and Eckardt, Andr\'e},
  journal = {Phys. Rev. A},
  volume = {98},
  issue = {6},
  pages = {063621},
  numpages = {6},
  year = {2018},
  month = {Dec},
  publisher = {American Physical Society},
  doi = {10.1103/PhysRevA.98.063621},
  url = {https://link.aps.org/doi/10.1103/PhysRevA.98.063621}
}

@article{raciunas_modified_2016,
  title = {Modified interactions in a {F}loquet topological system on a square lattice and their impact on a bosonic fractional {C}hern insulator state},
  author = {Ra\ifmmode \check{c}\else \v{c}\fi{}i\ifmmode \bar{u}\else \={u}\fi{}nas, Mantas and \ifmmode \check{Z}\else \v{Z}\fi{}labys, Giedrius and Eckardt, Andr\'e and Anisimovas, Egidijus},
  journal = {Phys. Rev. A},
  volume = {93},
  issue = {4},
  pages = {043618},
  numpages = {10},
  year = {2016},
  month = {Apr},
  publisher = {American Physical Society},
  doi = {10.1103/PhysRevA.93.043618},
  url = {https://link.aps.org/doi/10.1103/PhysRevA.93.043618}
}

@article{radu_quasiparticle_2008,
  title = {Quasi-Particle Properties from Tunneling in the $\nu = 5/2$ Fractional Quantum {H}all State},
  author = {Radu, Iuliana P. and Miller, J. B. and Marcus, C. M. and Kastner, M. A. and Pfeiffer, L. N. and West, K. W.},
  year = 2008,
  month = may,
  journal = {Science},
  volume = {320},
  number = {5878},
  pages = {899--902},
  publisher = {American Association for the Advancement of Science},
  doi = {10.1126/science.1157560},
  urldate = {2026-02-18}
}

@article{read_paired_2000,
  title = {Paired States of Fermions in Two Dimensions with Breaking of Parity and Time-Reversal Symmetries and the Fractional Quantum {{Hall}} Effect},
  author = {Read, N. and Green, Dmitry},
  year = 2000,
  month = apr,
  journal = {Phys. Rev. B},
  volume = {61},
  number = {15},
  pages = {10267--10297},
  publisher = {American Physical Society},
  doi = {10.1103/PhysRevB.61.10267},
  urldate = {2026-02-18}
}

@article{repellin_fractional_2020,
  title = {Fractional {C}hern insulators of few bosons in a box: {H}all plateaus from center-of-mass drifts and density profiles},
  author = {Repellin, C. and L\'eonard, J. and Goldman, N.},
  journal = {Phys. Rev. A},
  volume = {102},
  issue = {6},
  pages = {063316},
  numpages = {12},
  year = {2020},
  month = {Dec},
  publisher = {American Physical Society},
  doi = {10.1103/PhysRevA.102.063316},
  url = {https://link.aps.org/doi/10.1103/PhysRevA.102.063316}
}

@article{rubio-abadal_floquet_2020,
  title = {Floquet Prethermalization in a {Bose-Hubbard} System},
  author = {Rubio-Abadal, Antonio and Ippoliti, Matteo and Hollerith, Simon and Wei, David and Rui, Jun and Sondhi, S. L. and Khemani, Vedika and Gross, Christian and Bloch, Immanuel},
  journal = {Phys. Rev. X},
  volume = {10},
  issue = {2},
  pages = {021044},
  numpages = {14},
  year = {2020},
  month = {May},
  publisher = {American Physical Society},
  doi = {10.1103/PhysRevX.10.021044},
  url = {https://link.aps.org/doi/10.1103/PhysRevX.10.021044}
}

@article{sun_optimal_2020,
  title = {Optimal Frequency Window for {{Floquet}} Engineering in Optical Lattices},
  author = {Sun, Gaoyong and Eckardt, Andr{\'e}},
  year = 2020,
  month = mar,
  journal = {Phys. Rev. Res.},
  volume = {2},
  number = {1},
  pages = {013241},
  publisher = {American Physical Society},
  doi = {10.1103/PhysRevResearch.2.013241},
  urldate = {2026-03-10}
}

@article{tai_microscopy_2017,
  title = {Microscopy of the Interacting {{Harper}}--{{Hofstadter}} Model in the Two-Body Limit},
  author = {Tai, M. Eric and Lukin, Alexander and Rispoli, Matthew and Schittko, Robert and Menke, Tim and {Dan Borgnia} and Preiss, Philipp M. and Grusdt, Fabian and Kaufman, Adam M. and Greiner, Markus},
  year = 2017,
  month = jun,
  journal = {Nature},
  volume = {546},
  number = {7659},
  pages = {519--523},
  publisher = {Nature Publishing Group},
  issn = {1476-4687},
  doi = {10.1038/nature22811},
  urldate = {2026-03-10},
  copyright = {2017 Macmillan Publishers Limited, part of Springer Nature. All rights reserved.},
  langid = {english}
}

@article{umucalilar_bulk_2023,
  title = {Bulk density signatures of a lattice quasihole with very few particles},
  author = {Umucal\ifmmode \imath \else \i \fi{}lar, R. O.},
  journal = {Phys. Rev. A},
  volume = {108},
  issue = {6},
  pages = {L061302},
  numpages = {5},
  year = {2023},
  month = {Dec},
  publisher = {American Physical Society},
  doi = {10.1103/PhysRevA.108.L061302},
  url = {https://link.aps.org/doi/10.1103/PhysRevA.108.L061302}
}

@article{vendeiro_machinelearning_2022,
  title = {Machine-Learning-Accelerated {Bose-Einstein} Condensation},
  author = {Vendeiro, Zachary and Ramette, Joshua and Rudelis, Alyssa and Chong, Michelle and Sinclair, Josiah and Stewart, Luke and Urvoy, Alban and Vuleti{\'c}, Vladan},
  year = 2022,
  month = dec,
  journal = {Physical Review Research},
  volume = {4},
  number = {4},
  pages = {043216},
  publisher = {American Physical Society},
  doi = {10.1103/PhysRevResearch.4.043216},
  urldate = {2023-09-08}
}

@Article{wang_measurable_2022,
	title={{Measurable signatures of bosonic fractional {C}hern insulator states and their fractional excitations in a quantum-gas microscope}},
	author={Botao Wang and Xiao-Yu Dong and André Eckardt},
	journal={SciPost Phys.},
	volume={12},
	pages={095},
	year={2022},
	publisher={SciPost},
	doi={10.21468/SciPostPhys.12.3.095},
	url={https://scipost.org/10.21468/SciPostPhys.12.3.095},
}

@article{wang_realization_2024,
  title = {Realization of Fractional Quantum {{Hall}} State with Interacting Photons},
  author = {Wang, Can and Liu, Feng-Ming and Chen, Ming-Cheng and Chen, He and Zhao, Xian-He and Ying, Chong and Shang, Zhong-Xia and Wang, Jian-Wen and Huo, Yong-Heng and Peng, Cheng-Zhi and Zhu, Xiaobo and Lu, Chao-Yang and Pan, Jian-Wei},
  year = 2024,
  month = may,
  journal = {Science},
  volume = {384},
  number = {6695},
  pages = {579--584},
  publisher = {American Association for the Advancement of Science},
  doi = {10.1126/science.ado3912},
  urldate = {2026-02-18}
}

@misc{wei_10megahertz_2026,
      title={A 10 Megahertz Spatial Light Modulator}, 
      author={Xin Wei and Zeyang Li and Abhishek V. Karve and Adam L. Shaw and David I. Schuster and Jonathan Simon},
      year={2026},
      eprint={2601.08906},
      archivePrefix={arXiv},
      primaryClass={quant-ph},
      url={https://arxiv.org/abs/2601.08906}, 
}

@article{wen_topological_1990,
  title = {Topological Orders in Rigid States},
  author = {Wen, X. G.},
  year = 1990,
  month = feb,
  journal = {Int. J. Mod. Phys. B},
  volume = {04},
  number = {02},
  pages = {239--271},
  publisher = {World Scientific Publishing Co.},
  issn = {0217-9792},
  doi = {10.1142/S0217979290000139},
  urldate = {2026-02-18}
}

@article{wigley_fast_2016,
  title = {Fast Machine-Learning Online Optimization of Ultra-Cold-Atom Experiments},
  author = {Wigley, P. B. and Everitt, P. J. and {van den Hengel}, A. and Bastian, J. W. and Sooriyabandara, M. A. and McDonald, G. D. and Hardman, K. S. and Quinlivan, C. D. and Manju, P. and Kuhn, C. C. N. and Petersen, I. R. and Luiten, A. N. and Hope, J. J. and Robins, N. P. and Hush, M. R.},
  year = 2016,
  month = may,
  journal = {Sci Rep},
  volume = {6},
  number = {1},
  pages = {25890},
  publisher = {Nature Publishing Group},
  issn = {2045-2322},
  doi = {10.1038/srep25890},
  urldate = {2026-03-10},
  copyright = {2016 The Author(s)},
  langid = {english}
}

@article{willett_interference_2023,
  title = {Interference Measurements of Non-{A}belian $e/4$ \& {A}belian $e/2$ Quasiparticle Braiding},
  author = {Willett, R. L. and Shtengel, K. and Nayak, C. and Pfeiffer, L. N. and Chung, Y. J. and Peabody, M. L. and Baldwin, K. W. and West, K. W.},
  journal = {Phys. Rev. X},
  volume = {13},
  issue = {1},
  pages = {011028},
  numpages = {19},
  year = {2023},
  month = {Mar},
  publisher = {American Physical Society},
  doi = {10.1103/PhysRevX.13.011028},
  url = {https://link.aps.org/doi/10.1103/PhysRevX.13.011028}
}

@article{willett_magneticfieldtuned_2013,
  title = {Magnetic-Field-Tuned {Aharonov-Bohm} Oscillations and Evidence for Non-{A}belian Anyons at $\nu=5/2$},
  author = {Willett, R. L. and Nayak, C. and Shtengel, K. and Pfeiffer, L. N. and West, K. W.},
  year = 2013,
  month = oct,
  journal = {Phys. Rev. Lett.},
  volume = {111},
  number = {18},
  pages = {186401},
  publisher = {American Physical Society},
  doi = {10.1103/PhysRevLett.111.186401},
  urldate = {2026-03-11}
}

@article{willett_observation_1987,
  title = {Observation of an Even-Denominator Quantum Number in the Fractional Quantum {{Hall}} Effect},
  author = {Willett, R. and Eisenstein, J. P. and St{\"o}rmer, H. L. and Tsui, D. C. and Gossard, A. C. and English, J. H.},
  year = 1987,
  month = oct,
  journal = {Phys. Rev. Lett.},
  volume = {59},
  number = {15},
  pages = {1776--1779},
  publisher = {American Physical Society},
  doi = {10.1103/PhysRevLett.59.1776},
  urldate = {2026-02-18}
}

@article{wu_optimal_2025,
  title = {Optimal control for preparing fractional quantum {H}all states in optical lattices},
  author = {Wu, Ling-Na and Li, Xikun and Goldman, Nathan and Wang, Botao},
  journal = {Phys. Rev. B},
  volume = {111},
  issue = {23},
  pages = {235111},
  numpages = {11},
  year = {2025},
  month = {Jun},
  publisher = {American Physical Society},
  doi = {10.1103/PhysRevB.111.235111},
  url = {https://link.aps.org/doi/10.1103/PhysRevB.111.235111}
}

@article{zache_entanglement_2022,
  doi = {10.22331/q-2022-04-27-702},
  url = {https://doi.org/10.22331/q-2022-04-27-702},
  title = {Entanglement spectroscopy and probing the {L}i-{H}aldane conjecture in topological quantum matter},
  author = {Zache, Torsten V. and Kokail, Christian and Sundar, Bhuvanesh and Zoller, Peter},
  journal = {{Quantum}},
  issn = {2521-327X},
  publisher = {{Verein zur F{\"{o}}rderung des Open Access Publizierens in den Quantenwissenschaften}},
  volume = {6},
  pages = {702},
  month = apr,
  year = {2022}
}

@article{zupancic_ultraprecise_2016,
  title = {Ultra-Precise Holographic Beam Shaping for Microscopic Quantum Control},
  author = {Zupancic, Philip and Preiss, Philipp M. and Ma, Ruichao and Lukin, Alexander and Tai, M. Eric and Rispoli, Matthew and Islam, Rajibul and Greiner, Markus},
  year = 2016,
  month = jun,
  journal = {Opt. Express, OE},
  volume = {24},
  number = {13},
  pages = {13881--13893},
  publisher = {Optica Publishing Group},
  issn = {1094-4087},
  doi = {10.1364/OE.24.013881},
  urldate = {2026-03-10},
  copyright = {\copyright{} 2016 Optical Society of America},
  langid = {english}
}

\clearpage

\renewcommand{\thesubsection}{\Alph{subsection}}

\setcounter{figure}{0} 
\renewcommand{\figurename}{Extended Data FIG.}
\renewcommand{\thefigure}{\arabic{figure}}
\renewcommand{\theHfigure}{ExtendedData.\arabic{figure}} 

\begin{figure*}[t!!]
\centering
\includegraphics[width=\textwidth]{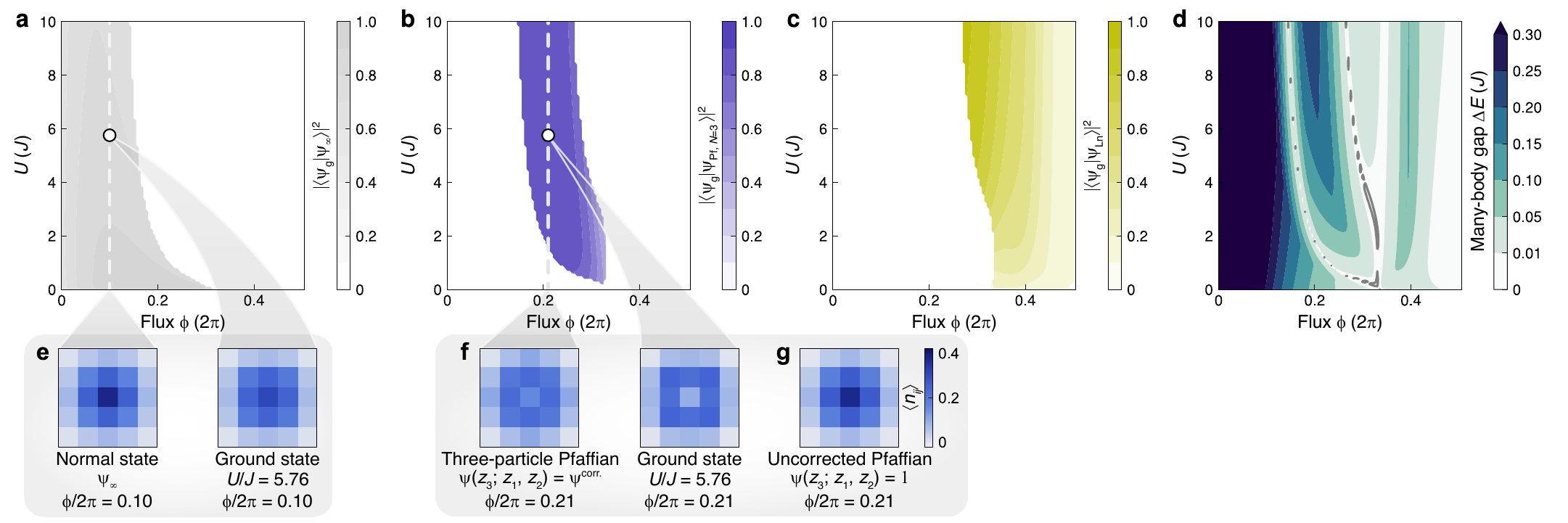}
\caption{
\textbf{Ground state of the Harper–Hofstadter Hamiltonian for three particles on a $\mathbf{5\times 5}$ lattice.}
\textbf{a-c}, Overlap of the ground state of the Harper–Hofstadter Hamiltonian at given parameters $U$ and $\phi$ with the analytic wave functions describing the normal (a), three-particle Pfaffian (equation~\eqref{eq:our_state_N=3}) (b), and Laughlin (c) states.
Together, the three plots form Fig.~\ref{fig:Intro}c of the main text.
White points indicate the parameters used for the ground-state densities shown below.
The contour lines are generated based on a data grid in steps of $\Delta U = 0.1 \ J$ and $\Delta \phi / 2\pi = 0.005$.
\textbf{d}, Many-body gap to the first excited state.
Where the gap closes around the ‘Pfaffian island' (gray dotted line), the ground state changes discontinuously, leading to the sharply defined areas of subplots a-c.
This allows the combined plotting of Fig.~\ref{fig:Intro}c without losing any information.
\textbf{e}, Comparing the density distributions of the analytical normal state wave function to the ground state at $\phi/2\pi = 0.10$. Both states feature the highest density in the center of the system.
\textbf{f}, By contrast, the three-particle Pfaffian wave function defined in equation~\eqref{eq:our_state_N=3} and the $\phi/2\pi = 0.21$ ground state feature a ring-shaped density profile.
\textbf{g}, This is not the case for the uncorrected Pfaffian wave function with the simplest possible choice $\psi(z_3; z_1, z_2) = 1$. The overlap of this wave function with the ground state is near zero for all considered values of $U$ and $\phi$.
}
\label{ext_data:analytic_overlaps}
\end{figure*}

\section*{Methods}
\label{sec:methods}

\begin{figure*}[t!!]
\centering
\includegraphics[width=\textwidth]{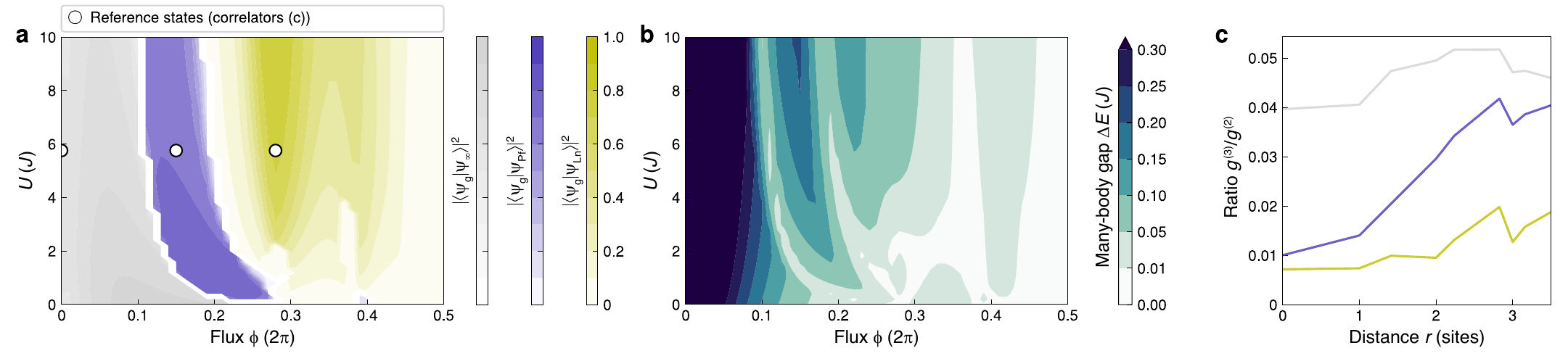}
\caption{
\textbf{Four particles on a $\mathbf{6 \times 6}$ lattice.}
As a comparison to the numerical results presented in Fig.~\ref{fig:Intro}c of the main text and Extended Data Fig.~\ref{ext_data:analytic_overlaps}, we present the corresponding results for a four-particle system on a larger lattice.
In this system, the analytical normal, Pfaffian, and Laughlin wave functions are defined without the modifications necessary for the odd particle number $N=3$ used in the main part of our work.
The $N=4$ system therefore provides an important reference for the robustness of our results with increasing system size.
\textbf{a-b}, The overlaps between the analytic wave functions and ground state at given $\phi$ and $U$, as well as the many-body gap, follow a structure very similar to our findings in the $N=3$ system.
In particular, we find high overlap with the Pfaffian state on an ‘island' at finite $U>0$, surrounded by a region where the gap becomes small and the ground state changes abruptly.
The contour lines are generated based on a data grid in steps of $\Delta U = 0.2 \ J$ and $\Delta \phi / 2\pi = 0.01$.
\textbf{c}, As in Fig.~\ref{fig:densities_and_correlations}d of the main text, the reduced three-body correlations $g^{(3)}/g^{(2)}$ in the ground state serve as a measure for distinguishing the different FQH states.
As in the three-particle system, the Pfaffian state exhibits the characteristic suppression of three-body correlations at short distances, while the normal and Laughlin regimes show consistently enhanced or suppressed correlations.
}
\label{ext_data:analytic_wave_functions}
\end{figure*}

\subsection{Analytic wave functions}
\label{sec:methods_analytic}
To choose the correct parameters for the on-site interaction $U$ and flux per plaquette $\phi$ in the experiment, we need to relate the analytical wave functions describing quantum Hall states to the ground state of the Harper–Hofstadter Hamiltonian.
We define the wave functions on our discrete $5 \times 5$ lattice by introducing the shifted complex coordinates
\begin{align}
    z_i = (x_i - x_0) + i(y_i - y_0),
\end{align}
where $x_i; \: y_i \in [1, 2, ..., 5]$ label the lattice sites and $(x_0, \: y_0) = (3, \: 3)$ is the center of the lattice.
This allows us to use the usual definition of the FQH wave functions~\cite{laughlin_anomalous_1983, moore_nonabelions_1991}:
\begin{align}
    \Psi_{\infty}[z] &= \prod_i^N \text{e}^{-\frac{1}{4 \ell_{\!B}^2}|z_i|^2}, \label{eq:Normal}\\
    \Psi_{1}^\mathrm{Pf}[z] &= \mathrm{Pf} \left(\frac{1}{z_i - z_j} \right)
     \prod_{i<j}^N(z_i - z_j) \: \Psi_{\infty}[z]. \label{eq:Pfaffian} \\
    \Psi_{1/2}^\mathrm{Ln}[z] &=
     \prod_{i<j}^N(z_i - z_j)^2 \: \Psi_{\infty}[z].
    \label{eq:Laughlin}
\end{align}
Here, $\Psi_{\infty}$ (the subscript indicates ${\nu}$) describes a droplet of independent particles, confined to one magnetic length $\ell_{\!B} = 1 / \sqrt{\phi}$.
In $\Psi_{1/2}^\mathrm{Ln}$, multiplication with repulsive factors $(z_i - z_j)^2$ between all the particles yields the $\nu = 1/2$ Laughlin state realized in previous work~\cite{leonard_realization_2023}.
To obtain the Pfaffian state $\Psi_{1}^\mathrm{Pf}$, the Laughlin wave function is multiplied by the Pfaffian factor 
\begin{align}
\mathrm{Pf} \left(\frac{1}{z_i - z_j} \right)\equiv \mathcal{A} \left\{\frac{1}{z_1 - z_2}\cdot\frac{1}{z_{3} - z_4}\ldots \right\}, 
\end{align}
where $\mathcal{A}$ denotes antisymmetrization.
The Pfaffian has three effects on the wave function, that is, on the Jastrow factor in equation~\eqref{eq:Pfaffian}. First, it pairs up particles. 
Second, it changes the symmetry under permutations, here from fermions to bosons. Third, it reduces the degree of the polynomial in each coordinate by 1, and hence the flux through the liquid by one flux quantum, $N_\Phi\to N_\Phi-1$.  This does not affect the filling factor $\nu$ defined by $\frac{1}{\nu}\equiv\frac{\partial N_\Phi}{\partial N}$, but can be important for small systems, as all particles are collectively drawn closer to the origin.

These definitions allow a direct comparison between the model wave functions for different regimes of $\phi$ and the ground states of the Harper–Hofstadter Hamiltonian depending on $U$ and $\phi$.
As illustrated in Fig.~\ref{fig:Intro}c (for $N=3$) and Extended Data Figs.~\ref{ext_data:analytic_overlaps} (for $N=3$) and~\ref{ext_data:analytic_wave_functions}a (for $N=4$), we find good overlaps between the low-$\phi$ ground states and the normal state $\Psi_{\infty}$ even for the strong interactions $U$ realized in the experiment.
At large fluxes, we find a significant overlap with the $\nu = 1/2$ Laughlin wave function $\Psi_{1/2}^\mathrm{Ln}[z]$.
In between these two cases, a third type of ground state emerges at finite interactions $U > 0$.
The region is characterized by filling factors
\begin{equation}
    \nu = \frac{N}{\frac{\phi}{2 \pi}(L - 1)^2} \approx 1,
\end{equation}
and therefore potentially hosts a Pfaffian state~\cite{Palm2021-BosonicPfaffianState}. 

For the odd particle number $N = 3$ realized in our work, one of the particles remains unpaired.
The extension of equation~\eqref{eq:Pfaffian} is
\begin{align}\
    \label{eq:N3_pfaffian}
    \Psi_{1}^{\mathrm{Pf},N=3}[z] 
    &= \mathcal{S} \{ \psi(z_3;z_1,z_2)(z_1-z_3)(z_2-z_3) \} \: \Psi_{ \infty}[z],
    \nonumber \\
\end{align}
where $\mathcal{S}$ denotes symmetrization and $\psi$ is a finite size correction.  This correction compensates the reduction of the total (canonical) total angular momentum due to the pairing, that is, due to the factor $\frac{1}{z_1-z_2}$.
As illustrated in Fig.~\ref{fig:Intro}c, we find very high overlaps with the ground state for
\begin{align}
    \label{eq:psi312}
    \psi^\mathrm{corr.}(z_3;z_1,z_2) = \frac{z_1 + z_2}{2} - z_3,
\end{align}
which increases the size of the droplet through a repulsion between the center-of-mass of the pair and the third particle.  Combining equation~\eqref{eq:psi312} with equation~\eqref{eq:N3_pfaffian} yields equation~\eqref{eq:our_state_N=3} in the main text.
Alternative choices for the finite size correction factor, such as 
$\psi^\mathrm{alt.}(z_3;z_1,z_2) = z_3$, result in good overlaps as well.

\begin{figure*}[t!!]
\centering
\includegraphics[width=\textwidth]{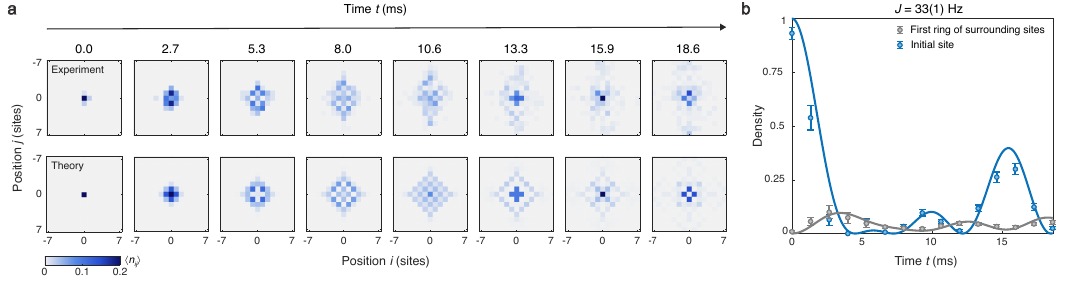}
\caption{
\textbf{Magnetic quantum walks at flux $\phi/2\pi = 0.21$}
\textbf{a}, Evolution of the single-particle quantum walk in 2D for experiment (top) and theory (bottom). 
\textbf{b}, The renormalized tunneling amplitude $K$ along $x$ is calibrated by fitting the densities on the initial site and on the first ring of surrounding sites as a function of time to the densities calculated by exact diagonalization. Error bars denote s.e.m.
}
\label{ext_data:MQW}
\end{figure*}

\subsection{Harper–Hofstadter Hamiltonian}
We use Floquet engineering to realize the Harper–Hofstadter Hamiltonian by periodically modulating the system parameters to simulate an effective Hamiltonian. In the laboratory frame, the system is described by
\begin{align}
    \hat{H}(t) =&
    - J_y \sum_{x, y} (\hat{a}^\dagger_{x, y+1} \hat{a}^{\vphantom{\dagger}}_{x, y} + \mathrm{H.c.})
    + \Delta_y \sum_{x, y} y \,\hat{n}_{x, y} \nonumber\\
    &- J_x \sum_{x, y} ( \hat{a}^\dagger_{x+1, y} \hat{a}^{\vphantom{\dagger}}_{x, y} + \mathrm{H.c.})
    + E_x \sum_{x, y} x \, \hat{n}_{x, y} \nonumber\\
    &+ \frac{U}{2} \sum_{x, y} \hat{n}_{x, y} (\hat{n}_{x, y} - 1) \nonumber\\
    &+ V_0 \sum_{x, y} \sin(\omega t + \phi_{x, y}) \,\hat{n}_{x, y}, 
    \label{eq:hamiltonian_lab_frame}
\end{align}
where $J_y$ and $\Delta_y$ denote the tunneling amplitude and energy tilt per site along $y$, while $J_x$ and $E_x$ denote the corresponding tunneling amplitude and tilt along $x$. The parameter $U$ is the on-site repulsive interaction energy, $V_0$ sets the strength of the periodic drive with angular frequency $\omega$. The site-dependent drive phase is $\phi_{x,y}=x\phi_x + y\phi_y$, where $\phi_x$ and $\phi_y$ are the phase difference between neighboring columns and rows, respectively. 

To obtain the Harper–Hofstadter Hamiltonian, we move to a rotating frame defined by the unitary transformation 
\begin{equation}
\hat{U}(t) = \exp\left[i\sum_{x,y}\left(-\frac{V_0}{\hbar\omega}\cos(\omega t + \phi_{x,y}) + x\omega t\right)\hat{n}_{x,y}\right],
\end{equation}
which removes the time-dependent on-site modulation. Then, applying the rotating-wave approximation yields the effective Hamiltonian
\begin{align}
    \hat{\mathcal{H}} =& - J \sum_{x, y} (\hat{a}^\dagger_{x, y+1} \hat{a}^{\vphantom{\dagger}}_{x, y} + \mathrm{H.c.})
    + \Delta_y \sum_{x, y} y \,\hat{n}_{x, y} \nonumber\\
    &- K \sum_{x, y} (e^{-i \phi_{x, y}} \hat{a}^\dagger_{x+1, y} \hat{a}^{\vphantom{\dagger}}_{x, y} + \mathrm{H.c.})
    + \Delta_x \sum_{x, y} x \, \hat{n}_{x, y} \nonumber\\
    &+ \frac{U}{2} \sum_{x, y} \hat{n}_{x, y} (\hat{n}_{x, y} - 1), \label{eq:hh_hamiltonian}
\end{align}
where $J \equiv J_y$ and $\Delta_x=E_x - \hbar\omega$. The tunneling amplitude along $x$ is renormalized to $K = J_x\mathcal{J}_1\left[\frac{V_0}{\hbar\omega}\sin(\frac{\phi_x}{2})\right]$, where $\mathcal{J}_1$ is the first-order Bessel function of the first kind, and its magnitude depends on the phase difference between adjacent columns. Crucially, the tunneling along $x$ acquires the complex phase $e^{-i\phi_{x,y}}$, set by the site-dependent modulation phase $\phi_{x,y}$; this phase acts as the Peierls phase associated with the synthetic magnetic field. Because $\phi_{x,y}$ changes by $\phi_y$ between adjacent rows, the effective flux through each plaquette is $\phi_y$, which we denote as $\phi$ in the main text. The drive restores tunneling when $\hbar\omega\approx E_x$; for slight detuning ($\hbar\omega< E_x$), a residual tilt $\Delta_x$ remains in the effective model.

\begin{figure*}[t!!]
\centering
\includegraphics[width=\textwidth]{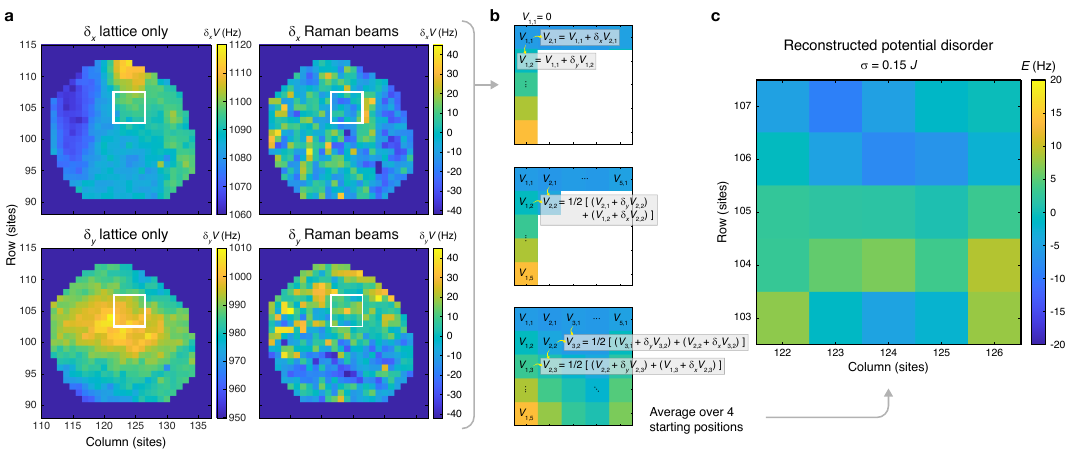}
\caption{
\textbf{Disorder measurements and local potential landscape.}
\textbf{a}, Gradient measurements used to reconstruct the potential in the region of interest. The top and bottom rows show the gradients along $x$ and $y$, respectively. In each row, the left panel shows the reference gradient (without Raman beams), and the right panel shows the Raman-beam gradient with the reference subtracted, isolating the disorder contribution from the individual Raman beams. The white squares mark the $5\times 5$ region of interest used in the experiment, chosen for its relatively low overall disorder (standard deviation $\sigma\approx 0.15~J$) and slight harmonic anti-confinement along $y$, which simulations show broadens the many-body gap to enable an adiabatic ramp to the Pfaffian state (Section~\ref{sec:interplayDisorder} and Extended Data Fig.~\ref{ext_data:rampNumerics}).
\textbf{b}, Schematic of the reconstruction of the potential disorder $V_{x,y}$ from the measured gradients $\delta_x V_{x,y}$ and $\delta_y V_{x,y}$. Starting from one corner (upper left in the example shown), the disorder is first calculated along the initial row and column, then extended to the rest of the system along the diagonal. This procedure is repeated for all four starting corners, and the resulting disorder patterns are averaged to estimate the disorder in the region of interest.
\textbf{c}, Reconstructed potential disorder in the experimental region of interest, obtained using the method described in Section~\ref{sec:EandUDisorder}. This disorder is important for understanding the optimized ramps, as shown in Extended Data Fig.~\ref{ext_data:rampNumerics}.
}
\label{ext_data:E_disorder}
\end{figure*}

\subsection{Calibration of Hamiltonian parameters}

\subsubsection{Tilts $\Delta_x$ and $\Delta_y$, on-site interaction $U$}
\label{sec:EandUCal}
We simultaneously calibrate the tilt $\Delta_x$ along $x$, generated by a magnetic-field gradient in that direction, and the on-site interaction energy $U$ using lattice-depth modulation spectroscopy~\cite{ma_photonassisted_2011}. Starting with an $n=1$ Mott insulator, we apply a magnetic-field gradient along $x$ and lower the lattice depth to $V_x = 15~E_\textrm{R}$ while keeping the lattice along $y$ deep at $V_y = 45~E_\textrm{R}$, restricting tunneling to the $x$ direction. Here $E_{\textrm{R}} = h\times 1.024$~kHz is the recoil energy for $^{87}$Rb atoms in a lattice with spacing $a=680$~nm. We then modulate the lattice depth $V_x$ across a frequency range and observe resonances at $\Delta_x - U'$ and $\Delta_x + U'$, where $U'$ is the on-site interaction energy at $V_x = 15~E_\textrm{R}$. At these resonances, atoms can tunnel onto already occupied sites. Because fluorescence imaging parity-projects the on-site occupation, light-assisted collisions remove atom pairs and thus reduce the probability of measuring odd occupation~\cite{bakr_quantum_2009}. We extract $\Delta_x$ and $U'$ from these resonances, and determine the on-site interaction energy in the actual experiment $U$ by the scaling relation
\begin{equation}
U = \left( \frac{V_{x,\textrm{exp}}}{V_{x,\textrm{cal}}} \right )^{1/4} \left(\frac{V_{y,\textrm{exp}}}{V_{y,\textrm{cal}}} \right)^{1/4} U',
\end{equation}
where $V_{x,\textrm{exp}} = 5.31~E_\textrm{R}$ and $V_{y,\textrm{exp}} = 11.483~E_\textrm{R}$ are the lattice depths used for the Pfaffian state, $V_{x,\textrm{cal}} = 15~E_\textrm{R}$ and $V_{y,\textrm{cal}} = 45~E_\textrm{R}$ are the lattice depths used in this calibration. Using the measured value $U' = 347(1)$~Hz, we obtain $U = 190.2(5)~\textrm{Hz} = 5.76(2)~J$. 

We calibrate the tilt $\Delta_y$ using an analogous procedure, now applying a magnetic-field gradient along $y$ and lowering the lattice depth along $y$ while keeping the lattice depth along $x$ deep.

\subsubsection{Flux $\phi$, tunneling amplitudes $J$ and $K$}
We realize synthetic magnetic fields using the approach described in~\cite{tai_microscopy_2017}, where a running-wave lattice is projected onto the atoms. To calibrate the flux per plaquette $\phi$, we set the frequencies of the two beams forming the running-wave lattice to be equal such that it becomes a standing wave. This standing wave is imprinted onto a superfluid from which we extract the wave vector $\textbf{k}$ by fitting the Fourier transform of the resulting interference fringes. The flux is then determined from $\phi/2\pi = \textbf{k}\cdot \hat{\textbf{e}}_y/ k_{\textrm{lat}}$, where $k_\textrm{lat} = 2\pi/a$ is the reciprocal lattice vector of the optical lattice with spacing $a$.

The tunneling amplitude $J$ along $y$ is calibrated by fitting the density profile after a single-particle quantum walk to $\rho_{|i|}(t) = |\mathcal{J}_i(2Jt)|^2$, where $\mathcal{J}_i$ is the Bessel function of the first kind and $i$ is the distance from the initial site~\cite{preiss_strongly_2015}. The tunneling amplitude $K$ along $x$ is calibrated in the same way, but with the tunneling restored by the running-wave lattice through Raman-assisted transitions. 

We then fine-tune both tunneling amplitudes using 2D single-particle quantum walks in the presence of a synthetic magnetic field, again with tunneling along $x$ restored by the running-wave lattice. The resulting density profiles show the decay and revival of the initial-site population, consistent with cyclotron motion in a magnetic field (Extended Data Fig.~\ref{ext_data:MQW}). By fitting the densities on the initial site and on the first ring of surrounding sites as a function of time to exact-diagonalization calculations, we obtain $J/h = K/h = 33(1)$~Hz. Because asymmetries in the magnetic quantum-walk densities indicate a residual tilt in the lattice, we also use these measurements to fine-tune the residual tilt.

\subsection{Spatially resolved potential disorder}
\label{sec:EandUDisorder}
The modulation-spectroscopy technique used to calibrate the tilts $\Delta_x$ and $\Delta_y$ and the on-site interaction energy $U$ described in Section~\ref{sec:EandUCal} can also be used to estimate the disorder introduced by the Raman beams at the single-site level. In the standard calibration, resonant photon-assisted tunneling produces a sharp drop in the probability of measuring odd occupation, which is averaged over a large region to suppress local disorder and isolate the global tilt along a given axis. With sufficient data, the same method can instead be applied locally to measure the spatial variation of the potential gradient across the Mott insulator. These local gradients are then used to reconstruct an approximate disorder landscape in the experimental region of interest.

The spatial resolution of this method is limited by parity projection in the imaging. Near resonance with the local tilt shifted by the interaction energy, that is,  $\Delta_{\mathrm{loc}}+U$ ($\Delta_{\mathrm{loc}}-U$), the probability of even occupation increases, which manifests as a reduction in odd occupation due to parity projection. On a given site $j$, however, this signal combines two indistinguishable processes: loss of the initial atom as it tunnels uphill (downhill) from site $j$ to $j+1$ ($j-1$), and gain of an additional atom that has tunneled up (down) from $j-1$ ($j+1$). As a result, the measured response at site $j$ reflects the average of the potential gradients on the two adjacent links, $j$ to $j+1$ and $j$ to $j-1$.

Reconstructing the full Raman beam disorder profile requires six different local gradient measurements, two measurements (to measure the gradient along $x$ and along $y$) for each of three cases: a reference measurement without the Raman beams and a measurement with each beam separately (this separation is necessary to eliminate the sinusoidal interference and measure only the potential offset from variations in beam intensity). 
Subtracting the reference measurement from the corresponding Raman-beam measurement yields the local gradients $\delta_x V_{x,y}$ and $\delta_y V_{x,y}$, which serve as a discrete partial derivatives of the potential and are used to reconstruct the disorder landscape in the experimental region of interest (Extended Data Fig.~\ref{ext_data:E_disorder}a).

At this point, the uphill/downhill ambiguity once again comes into play, and reconstructing the local potential $V_{x,y}$ from the measured gradients is not unique because the parity-projected signal assigns each site the average gradient of its two neighboring links. As a result, there are four equivalent ways to reconstruct the potential landscape, corresponding to the four possible choices of starting corner in the region of interest (Extended Data Fig.~\ref{ext_data:E_disorder}b). In each case, the potential at the starting corner is set to zero and the potential on the remaining sites is obtained recursively by summing the measured gradients along the $x$ and $y$ directions. For an arbitrary site $(x,y)$, the four reconstructions are
\begin{align*}
V^{(1)}_{x,y} = \frac{1}{2} &\left(V^{(1)}_{x-1,y} + \delta_x V_{x,y} + V^{(1)}_{x,y-1} + \delta_y V_{x,y} \right), \\ 
V^{(2)}_{x,y} = \frac{1}{2} &\left(V^{(2)}_{x-1,y} + \delta_x V_{x,y} + V^{(2)}_{x,y+1} - \delta_y V_{x,y}\right), \\
V^{(3)}_{x,y} = \frac{1}{2} &\left(V^{(3)}_{x+1, y} - \delta_x V_{x,y} + V^{(3)}_{x,y-1} + \delta_y V_{x,y}\right), &\textrm{and} \\
V^{(4)}_{x,y} = \frac{1}{2} &\left(V^{(4)}_{x+1, y} - \delta_x V_{x,y} + V^{(4)}_{x,y+1} - \delta_y V_{x,y}\right).
\end{align*}
On the edges of the region of interest, where one of the two neighboring links is absent, the potential is updated using the available single-axis contribution only. The reconstructed disorder landscape is then obtained by averaging these four cases.

This averaging smooths the potential across neighboring sites and may therefore underestimate the true disorder strength. Even with this limitation, the reconstructed disorder landscape captures the experimentally relevant structure well enough to reproduce several key observations numerically, most notably the adiabaticity of the experimentally optimized ramp (Extended Data Fig.~\ref{ext_data:rampNumerics}). The agreement is nevertheless limited by additional residual tilts that are not included in the disorder reconstruction. In the experiment, the overall gradients along $x$ and $y$ are tuned in situ to center the atoms within the region of interest. This leaves an offset tilt in each direction as an additional free parameter, which is not captured by the disorder measurement but can still significantly affect the density profile in numerical simulation.

\subsection{Engineered light potentials}
We holographically project engineered light potentials at 760~nm using two digital micromirror devices (DMDs). The DMDs allow us to project arbitrary potentials while correcting optical aberrations in the imaging system to achieve diffraction-limited performance~\cite{zupancic_ultraprecise_2016}. We use the DMDs both to prepare the initial state and to project confinement during the ramp to the target state. 

To prepare the initial state, we use a cookie-cutting protocol to prepare the $\ket{1,1,1,0,0}$ Fock state with high fidelity. The cutting is performed sequentially along the $x$ and $y$ directions. In the first step, we isolate a single column of atoms from the $n=1$ region of the Mott insulator using a $\textrm{TEM}_{10}$-like potential. In the second step, we isolate three adjacent atoms along $y$ using a $\textrm{TEM}_{03}$-like potential. After each step, the remaining atoms are removed by applying a repulsive Gaussian beam while lowering the corresponding lattice depth, $V_x$ or $V_y$, to zero. The lattice depth is then restored to $45~E_{\textrm{R}}$ in preparation for ramping to the target state.

The DMDs are subsequently used to project repulsive walls along both $x$ and $y$. Each wall has a Gaussian profile with a width $w_0=0.85$~sites and is placed 1.5~sites from the edge of the $5 \times 5$ system. The wall height is chosen to be approximately 800~Hz, large enough to constrain the low-energy dynamics of the system while avoiding resonance with the Raman-drive frequency. The disorder introduced by the walls to the $5\times 5$ system is $\leq 1$~Hz, negligible compared to the tunneling energy. 

\begin{figure}[t!!]
\centering
\includegraphics[width=\linewidth]{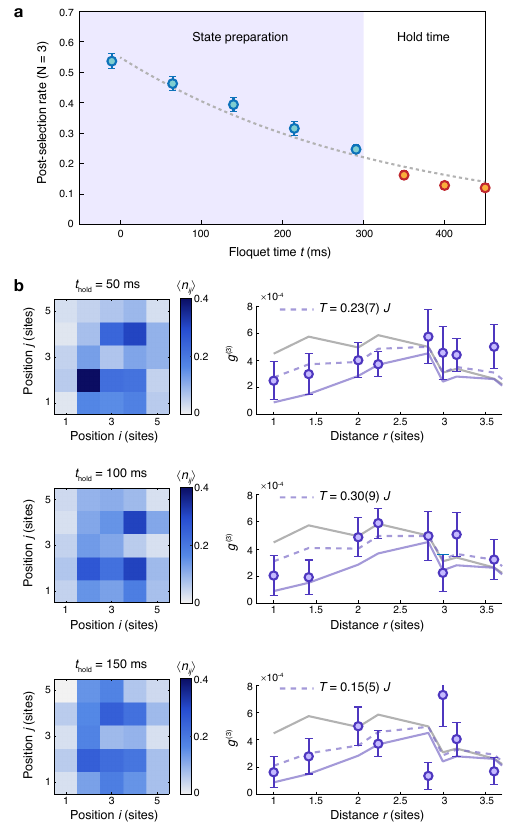}
\caption{
\textbf{Single-particle lifetime and coherence time under the Floquet drive.}
\textbf{a}, Post-selection rate for $N = 3$ measured at different times under the Floquet drive. The shaded region marks different points during state preparation when the Floquet drive is active. The unshaded region shows measurements with an additional hold time $t_{\mathrm{hold}}$ after state preparation.
The $x$-axis is the total time with which the system is subject to the Floquet drive. Assuming density-independent particle loss, the post-selection rate is fit to $P(t) = A\exp(-3t/\tau_{\mathrm{single}})$, yielding a single-particle lifetime of $\tau = 1.0(3)$~s. Error bars denote s.e.m.  
\textbf{b}, Densities and $g^{(3)}_{\mathbf{i}, \mathbf{i'}, \mathbf{j}} = \langle \hat{n}_\mathbf{i} \hat{n}_\mathbf{i'} \hat{n}_\mathbf{j} \rangle$ correlations after a hold time $t_{\mathrm{hold}}$. Neither shows a significant change up to the longest measured hold time of $150$~ms, although the post-selection rate limits measurements at longer times. Each hold-time measurement consists of $\sim 130$ snapshots. The fitted temperatures of the correlations (dashed purple) are still consistent with the state preparation fidelity, suggesting that the many-body coherence time is at least $500$~ms ($\sim 100~\tau$), sufficient for state-preparation and the Hall drift measurement. Error bars denote bootstrap-estimated 68\% confidence intervals.
}
\label{ext_data:coherence_time}
\end{figure}

\subsection{Optimal Floquet regime}
\label{sec:optFloquetRegime}
Implementing the synthetic magnetic field with a Raman drive subjects the system to Floquet heating. To minimize this, we operate in an optimal Floquet regime and choose the drive parameters according to the principle in~\cite{sun_optimal_2020}, namely by separating the relevant energy scales and avoiding resonances. Driving at too high a frequency can couple atoms to higher bands and induce atom loss or intraband heating~\cite{rubio-abadal_floquet_2020}, whereas driving at too low a frequency enhances higher-order terms in the effective Hamiltonian that can destabilize topological states~\cite{raciunas_modified_2016,hudomal_bosonic_2019}.

In our experiment, the high-energy scale is set by the band gap between the lowest and first excited bands, approximately 3~kHz for $V_x = 5.31~E_\textrm{R}$ and 5~kHz for $V_y = 11.48~E_\textrm{R}$, the lattice depths used to realize the Pfaffian state. The low-energy scale is set by the bare tunneling $J_x/h = 84(1)$~Hz and the interaction strength $U = 190(1)$~Hz. We choose a driving frequency $\omega/2\pi = f_{\textrm{Raman}} = 1.1$~kHz, giving a ratio $f_{\textrm{Raman}} / (J_x/h) \approx 13$. This ratio places the system in the high-frequency limit with respect to the Hubbard parameters while remaining off resonance with both the interaction scale and the higher bands. These Floquet parameters give an estimated single-particle lifetime of $1.0(3)$~s and a coherence time $\geq 500$~ms for the interacting three-particle system (Extended Data Fig.~\ref{ext_data:coherence_time}).

\begin{figure*}[p!]
\centering
\includegraphics[width=\textwidth]{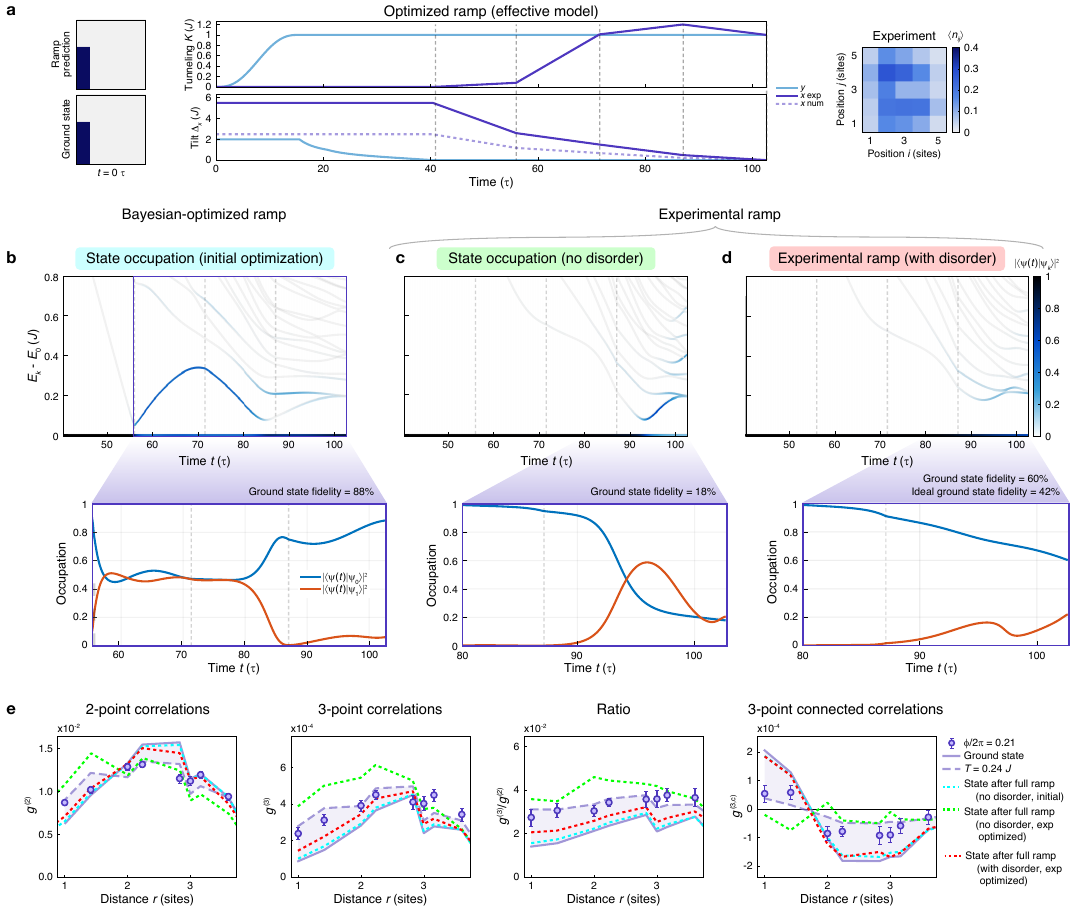}
\caption{
\textbf{Role of disorder in the ramp preparation.} After Bayesian optimization in an ideal system, we tune experimental parameters (the initial tilt along $x$) to optimize the return overlap in the experiment, obtaining a ramp that rescales the numerically optimized one. The tilt and tunneling in both cases are shown in
\textbf{a}, with the original tilt ramp indicated by the dashed line. Using the measured disorder (see Section~\ref{sec:EandUDisorder} and Extended Data Fig.~\ref{ext_data:E_disorder}), we numerically simulate the full experimental ramp and interpret the change as a response to disorder.
\textbf{b}, Upper, energy gap versus time during the final stage of the preparation, when tilt and tunneling are ramped simultaneously (steps 3a and 3b in Fig.~\ref{fig:Expsequence}a of the main text). The shading of each level indicates the time-dependent occupation. For the initial optimized ramp, the evolution is diabatic: the occupation splits into the first excited state and later recombines, yielding an expected overlap of 88\% in the ideal system. Lower, expanded view near the end of the ramp showing the occupations of the ground and first excited states versus time.
\textbf{c}, Same numerical analysis for the ramp used in the experiment, after optimization of the initial tilt along $x$ (see Section~\ref{sec:interplayDisorder}). In the simulation, the gap closes once near the final step of the ramp, and the ground-state occupation drops to 18\%.
\textbf{d}, Same ramp as in c, but now including the measured disorder. The disorder widens the gap to the first excited state near the end of the ramp, allowing adiabatic evolution and yielding a final fidelity of 60\% with respect to the disordered ground state. This corresponds to a fidelity of 42\% relative to the ideal, disorder-free ground state (the ideal and disordered ground states overlap by 81\%).
\textbf{e}, Correlations versus distance for several cases. The data points are the same as in Fig.~\ref{fig:densities_and_correlations} of the main text. The solid purple and light dashed purple lines show the ideal ground-state and finite-temperature correlations, respectively. The blue dashed line shows the expected final state from exact-diagonalization simulations of the numerically optimized ramp, which deviates only slightly from the ideal case. The green dashed line shows the numerical correlations for the ramp used in the experiment; these lie much farther from the ideal case than the experimental data and outside the range accounted for by finite temperature. The red dashed line shows the numerical correlations for the same ramp including the experimentally estimated potential disorder, recovering agreement with the ideal case. Error bars denote bootstrap-estimated 68\% confidence intervals.
}
\label{ext_data:rampNumerics}
\end{figure*}

\subsection{Optimized state preparation}
\label{sec:State Preparation}
We use Bayesian optimization to design the final stage of state preparation, where atoms in the 1D superfluid are delocalized along the $x$ direction to prepare the 2D Pfaffian state (steps 3a and 3b in Fig.~\ref{fig:Expsequence}a of the main text).
Following the strategy outlined in~\cite{blatz_bayesian_2024}, we first optimize the state-preparation sequence using exact-diagonalization simulations and then refine the resulting protocol by calibrating key parameters in the experiment.

\subsubsection{Bayesian optimization}
For each iteration of the optimization, we numerically simulate the evolution from the
initial state, where the three particles are localized in the bottom-left corner of the lattice (step 1 in Fig.~\ref{fig:Expsequence}a), at time $t = 0$ to the final state at $t = t_f \approx 100~\tau$ under the Harper–Hofstadter Hamiltonian in equation~\eqref{eq:hh_hamiltonian} with time-dependent parameters $J(t)$, $K(t)$, $\Delta_y(t)$ and $\Delta_x(t)$.
First, from $t = 0$ to $t = t_\text{1D}$, the system is delocalized into a one-dimensional superfluid (step 2 in Fig.~\ref{fig:Expsequence}a) using a fixed local-adiabatic protocol. In this regime, the many-body gap is large, so a duration of $t_\text{1D} \approx 40~\tau$ is sufficient for adiabatic preparation, during which $J(t)$ and $\Delta_y(t)$ are sequentially ramped to their final values (see Extended Data Fig.~\ref{ext_data:rampNumerics}).
Bayesian optimization is then applied to the ramps of $K(t)$ and $\Delta_x(t)$, which control delocalization along the $x$ direction from $t = t_\text{1D}$ to $t_f = t_\text{1D} + t_\text{2D}$. Both $K(t)$ and $\Delta_x(t)$ are parametrized by four linear segments of equal duration and ramped simultaneously over $t_\text{2D} \approx 60~\tau$ (steps 3a and 3b in Fig.~\ref{fig:Expsequence}a). With fixed initial and final values, this leaves six control parameters that specify the values $K(t_i)$ and $\Delta_x(t_i)$ at three intermediate times $t_i$.

We use the expectation value $\langle \Psi(t_f) \vert \hat{\mathcal{H}}(t_f) \vert \Psi(t_f) \rangle$ of the final state as the cost function, rather than the preparation fidelity $|\langle \Psi_0 | \Psi(t_f) \rangle|^2$, where $\Psi(t_f)$ is the prepared state at $t=t_f$ and $\Psi_0$ is the ground state of $\hat{\mathcal{H}}(t_f)$.
Typically, we find this choice of the cost function to lead to faster convergence with the number of iterations and to reduce the dependence on the initialization of the optimizer.
This behavior is somewhat intuitive, as the energy provides quantitative feedback on the preparation quality, even when the ground-state overlap is low.
The optimization is performed for a total of $400$ iterations, although in practice fewer than $100$ are typically needed for convergence.
Additionally, we find that including a fixed disorder pattern in the numerical simulation results in protocols that are more robust to unknown disorder than those optimized in the clean system.

\subsubsection{Calibrating $\Delta_x(t=0)$ and $t_{2\text{D}}$}
In the next step, the suggested ramp is implemented in the experiment, and the following two key parameters are chosen for manual calibration:
\begin{enumerate}
\item Global scaling of the tilt ramp $\Delta_x(t) \rightarrow \gamma \Delta_x(t)$.
As can be seen in Fig.~\ref{fig:Expsequence}b, this parameter controls whether the ramp passes the gap closing near the $(K, \: \Delta_x) = (0, \: 1.2~J)$ point, making it a switch between diabatic and adiabatic preparation.
\item When the preparation protocol is diabatic, the total ramp time $t_\text{2D}$ becomes an important parameter:
While the preparation fidelity is expected to increase monotonically with $t_\text{2D}$ for an adiabatic protocol, we find it to be peaked around some optimum time $t_\text{2D}^{(\mathrm{opt})}$ in simulations of the diabatic case.
\end{enumerate}

In the experiment, we scan the two parameters separately, and determine their optimum via the return probability (the probability to return to the initial state when inverting the preparation protocol), and measurements of the final state, such as its density distribution.
While we find the optimal preparation time to be close to the fixed time used for optimization,
manual calibration introduces a large scaling factor $\gamma \approx 2$, shifting the protocol from the diabatic to the adiabatic regime.

\subsubsection{Interplay with disorder}
\label{sec:interplayDisorder}
The fact that the calibration procedure described above produces a non-trivial factor $\gamma \neq 1$ demonstrates the importance of optimizing these final two parameters in the experiment, given their dependence on disorder.
As shown in Fig.~\ref{fig:Expsequence}b, crossing into the Pfaffian regime requires passing a region where the many-body gap reaches a minimum at $\Delta E \approx 0.1 \ J$, setting the timescale for adiabatic preparation.
This smallest gap determines whether adiabatic preparation is feasible within the coherence time of the experiment, which is limited by Floquet-induced heating of the Raman drive. Its size also depends sensitively on the disorder.
For the disorder profile measured in our region of interest (Extended Data Fig.~\ref{ext_data:E_disorder}c), the exact-diagonalization simulation predicts the gap to be raised significantly (Extended Data Fig.~\ref{ext_data:rampNumerics}b).
This lifting of the gap due to disorder explains why the adiabatic ramp obtained during manual calibration can realize the Pfaffian state in a sufficiently short time. We confirm this finding by repeating the optimization, including the measured disorder, which predicts protocols similar to the adiabatic ramp found by manual calibration. \\

The ability to adapt to an unknown disorder profile and switch between diabatic and adiabatic preparation, based on just two parameters, highlights the flexibility of the described state-preparation strategy. Extending the protocol to give direct experimental feedback to the Bayesian optimizer is a promising direction for future work.

\subsection{Post-selection}
During fluorescence imaging, inelastic light-assisted collisions induce pairwise atom loss such that the detected occupation reflects the parity of the initial atom number. Sites with even occupation appear empty, while sites with odd occupation appear singly occupied, preventing direct detection of multiply occupied sites. We therefore post-select images containing three atoms located on distinct lattice sites.

Given this criterion, the post-selection rates at successive stages of the experiment are as follows. (1) Using the DMD, we prepare an initial state of three atoms on adjacent sites from the $n=1$ region of the Mott insulator, yielding a post-selection rate of $78(2)\%$ and a preparation fidelity of $99.3(5)\%$, defined as the fraction of post-selected images with atoms located on the desired initial sites. (2) After performing the ramp that prepares the Pfaffian (normal) state, the post-selection rate decreases to $26.3(6)\%$ ($23.2(5)\%$), with resonant processes induced by the Floquet drive as the primary source of loss (Extended Data Fig.~\ref{ext_data:coherence_time}a).

After applying this post-selection criterion, the Pfaffian dataset used in the main text consists of 2,241 images. For comparison, the normal dataset at zero flux consists of 1,736 post-selected images.

Number states containing doubly or triply occupied sites can be neglected in the analysis without affecting the results. For the Pfaffian (normal) state at $T=0~J$, simulations yield probabilities of $1.8 \times 10^{-2}$ ($4.0 \times 10^{-2}$) for doubly occupied sites and $2.8 \times 10^{-5}$ ($8.6 \times 10^{-5}$) for triply occupied sites upon projection. At finite temperature these probabilities remain comparably small. Because such events are rare, we compare the post-selected experimental data directly with the full theoretical distribution, which includes states with multiply occupied sites, without introducing significant bias.

\begin{figure*}[t!!]
\centering
\includegraphics[width=\textwidth]{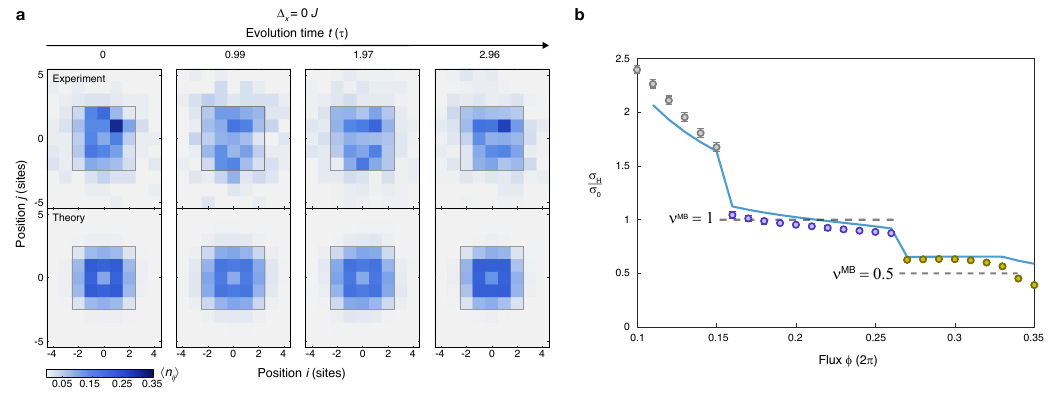}
\caption{
\textbf{The $\Delta_x = 0$ Hall drift measurement and numerical evidence for the Hall conductivity plateaus.} 
\textbf{a}, Full dataset of the $\Delta_x = 0$ Hall drift measurement. 
\textbf{b}, The Hall conductivity plateaus from numerical simulation. The simulation is performed using exact diagonalization for an $N=3$, $7 \times 11$ system. The ground state is released from the $5 \times 5$ system into the larger system after a quench of the confinement and applied force. The wall height and quench time are the same as in the experiment, and the drift velocity is extracted after an evolution of $5~\tau$ to reduce the effect of density oscillations (Section~\ref{sec:Halldrift}). The blue solid line shows the bulk filling factor $\nu_{\mathrm{bulk}}$ of the ground state. In the normal regime (gray), the Hall conductivity exhibits $\sim1/\phi$ scaling. In the Pfaffian (purple) and Laughlin (green) regimes, the Hall conductivity forms plateaus separated by abrupt jumps. The average plateau value in the Pfaffian regime is $\sigma_\mathrm{H} / \sigma_0 \approx 0.94$, in good agreement with the many-body Chern number $\nu^{\mathrm{MB}}=1$. In the Laughlin regime, the average plateau value is $\sigma_\mathrm{H} / \sigma_0 \approx 0.60$, close to the many-body Chern number $\nu^{\mathrm{MB}}=0.5$. Error bars denote the 68\% confidence interval from fitting the simulated center-of-mass motion.
}
\label{ext_data:hall_drift}
\end{figure*}

\subsection{Temperature fit}
We extract the temperature by comparing the measured probability distribution with simulated thermal ensembles of the many-body eigenstates of the Hamiltonian. The temperature is obtained by minimizing the Kullback–Leibler (KL) divergence 
\begin{equation}
D_{\mathrm{KL}}(P_{\mathrm{exp}}||P_T)=\sum_i P_{\mathrm{exp}}(i)\log\left[\frac{P_{\mathrm{exp}}(i)}{P_T(i)} \right],
\end{equation}
between the experimental distribution $P_{\mathrm{exp}}$ and the Boltzmann-weighted theoretical distribution $P_T$, where $i$ labels the many-body configurations in the number basis.

To estimate the uncertainty, we employ bootstrap resampling. For each bootstrap sample, we minimize the KL divergence to obtain a best-fit temperature. The reported temperature and uncertainty are the mean and standard deviation of the resulting distribution of temperatures. Using $1000$ bootstrap realizations, we obtain $T=0.24(1)~J$ for the Pfaffian dataset and $T=0.52(2)~J$ for the normal dataset.

\subsection{The Hall drift experiment}
\label{sec:Halldrift}
The Hall conductivity is extracted through the formula 
\begin{equation}
\sigma_\mathrm{H} / \sigma_0 = 2\pi\rho_{\mathrm{bulk}}v_{\perp}/F,
\end{equation}
where $F\equiv \Delta_x$ is the applied force in units of $J/\textrm{site}$~\cite{repellin_fractional_2020}. The Hall conductivity of the $N=3$, $5\times5$ system exhibits clear plateaus when the state is released into a larger system, as shown by the numerical simulation in Extended Data Fig.~\ref{ext_data:hall_drift}b. In the normal regime, the Hall conductivity $\sigma_\mathrm{H} / \sigma_0$ scales as $1/\phi$, whereas it forms plateaus near $0.94$ and $0.60$ for the $\nu=1$ and $\nu=1/2$ quantum Hall states, respectively.

In Fig.~\ref{fig:hall_drift}c of the main text, the drift velocity is extracted from a linear fit, $y = p_1 x + p_2$, to the $y$ center-of-mass drift. The oscillations in $\Delta Y_{c.o.m.}$ predicted by ground-state theory arise from coherent non-adiabatic release at a well-defined phase. In contrast, the experimental data show little visible oscillation, possibly due to finite preparation fidelity and fluctuations of the initial phase. Despite this, the measured drift velocity agrees well with theory at long evolution times ($\gtrsim 5~\tau$), when the coherent oscillations have been averaged out. Extended Data Fig.~\ref{ext_data:hall_drift}a shows the full dataset for $\Delta_x = 0$.

\subsection{Numerical simulations}
All theoretical predictions in the main text were obtained from exact diagonalization of the interacting Harper–Hofstadter Hamiltonian using the measured interaction strength $U=5.76~J$ and equal tunneling amplitudes along $x$ and $y$ ($K=J$), without free parameters. For the results in Figs.~\ref{fig:Intro},~\ref{fig:Expsequence},~\ref{fig:densities_and_correlations}, and \ref{fig:binning}, we performed exact diagonalization for a system size of three atoms on a $5 \times 5$ lattice. For Fig.~\ref{fig:hall_drift}, we simulated the time evolution using Krylov propagation on a $9 \times 11$ lattice. None of the simulations in the main text include disorder.  

\clearpage

\begin{figure*}[h!!]
\centering
\includegraphics[width=\textwidth]{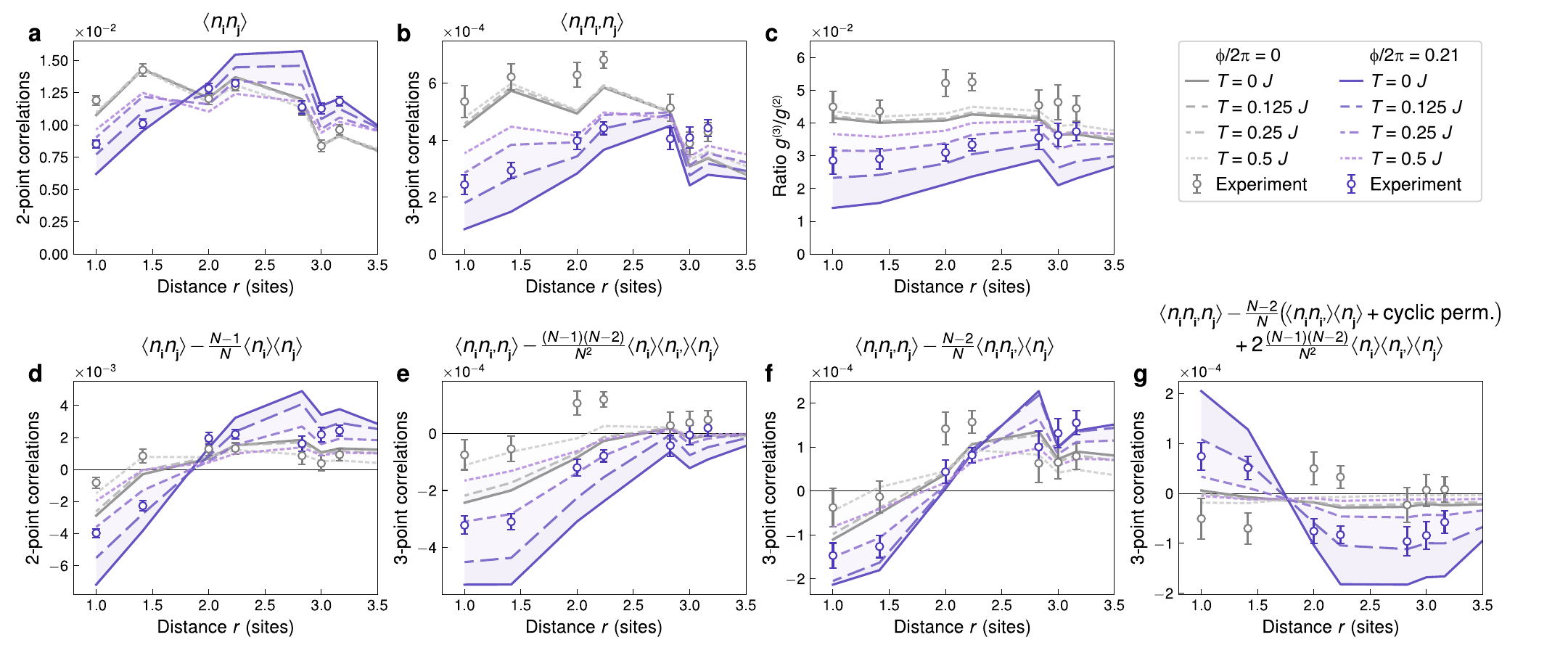}
\caption{
\textbf{Temperature dependence of correlation functions.}
\textbf{top}, Temperature dependence of the correlation functions presented in Fig.~\ref{fig:densities_and_correlations}.
\textbf{a-b}, As discussed in the main text, the data in the Pfaffian regime agree well with theory at $T \lesssim 0.24~J$.
The low temperature realized in our experiment is a prerequisite for distinguishing the paired state from the $\phi/2\pi = 0$ normal state within the experimental error bars.
\textbf{c}, Especially for the ratio $g^{(3)} / g^{(2)}$, the signal is predicted to become very weak at $T \gtrsim 0.5~J$.
\textbf{bottom}, This trend continues for the connected correlation functions, which serve as a sensitive test of the pairing physics.
\textbf{d}, Even the connected two-point correlations are predicted to become nearly indistinguishable from the $\phi/2\pi = 0$ normal state at $T \approx 0.5~J$. Due to the repulsive interactions $U > 0$, the normal state also shows anticorrelation at short distances.
\textbf{e}, Connected three-point correlations, comparing the observed correlations $\langle n_\mathbf{i} n_\mathbf{i'} n_\mathbf{j} \rangle$ to the uncorrelated density distribution.
The introduction of a synthetic magnetic field leads to suppressed correlations up to extended distances $d \sim 2$ compared to the $\phi/2\pi = 0$ reference.
\textbf{f}, Partly connected correlation function, subtracting the unconnected contribution of finding a pair of particles on nearest-neighbor sites $\mathbf{i}$ and $\mathbf{i'}$, and the third particle on site $\mathbf{j}$, making this definition a sensitive probe of the Pfaffian pairing.
\textbf{g}, Fully connected three-body correlations, subtracting all unconnected one- and two-point contributions.
In agreement with theory, we observe finite three-body correlations in the Pfaffian regime, which are expected to be near-zero in the normal state.
In summary, the experimental data agree well with theory at $T \lesssim 0.24~J$ for all correlation measures considered.
Comparison with theoretical predictions at different temperatures shows that low temperature is essential for resolving the features of the Pfaffian state. Error bars denote bootstrap-estimated 68\% confidence intervals.
}
\label{ext_data:connected_correlators}
\end{figure*}

\begin{figure*}[h!!]
\centering
\includegraphics[width=\textwidth]{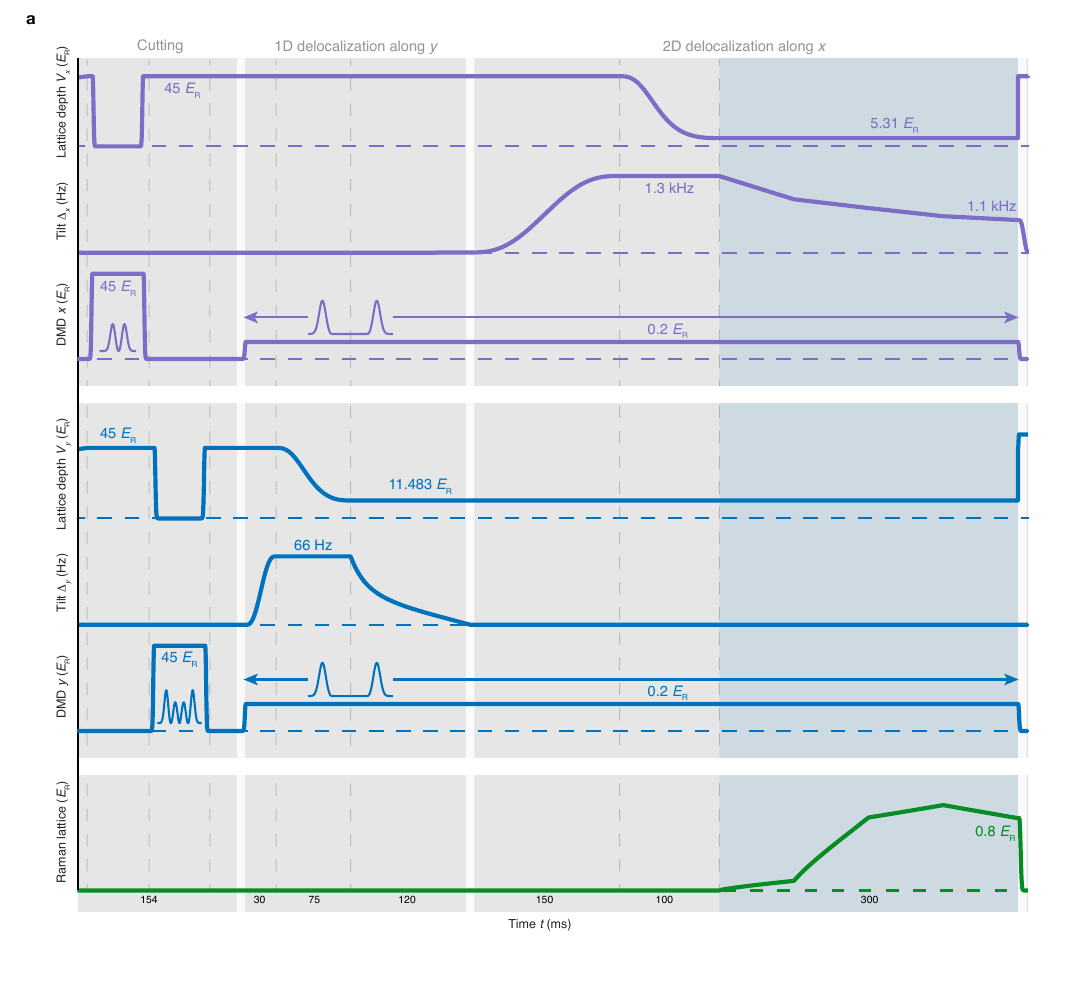}
\caption{
\textbf{Experimental sequence.} 
\textbf{a}, Full experimental sequence for key parameters tilt and tunneling along $x$ and $y$, and confinement of the system, from initializing the state $|1, 1, 1, 0, 0\rangle$ to preparing the Pfaffian state. All ramp values are shown within the full model in equation~\eqref{eq:hamiltonian_lab_frame}, with the relevant parameters plotted as a function of time. The final 300~ms (highlighted in blue) shows the optimized simultaneous ramp of Raman power (approximately proportional to $K$) and tilt along $x$ (proportional to residual gradient $\Delta_x$). This final segment of the tilt ramp along $x$ is not drawn to scale relative to the rest of the sequence, and is exaggerated for visual clarity.
}
\label{ext_data:ramps}
\end{figure*}

\end{document}